\documentclass[useAMS,usenatbib,doublespacing,referee]{mn2e}
\usepackage{graphicx}
\usepackage{epsfig}
\usepackage{color}

\title[CAB star: CF Tuc]{The chromospherically--active binary CF Tuc revisited}
\author[D. Do\u{g}ru, A. Erdem, S. S. Do\u{g}ru, S. Zola]{D. Do\u{g}ru$^{{1},{2}}$\footnotemark[1],
 A. Erdem$^{{1},{2}}$, S. S. Do\u{g}ru$^{{1},{2}}$, S. Zola$^{{3},{4}}$ \\
 \thanks{E-mail: dsurgit@comu.edu.tr (DD); aerdem@comu.edu.tr (AE); dogru@comu.edu.tr (SD); szola@oa.uj.edu.pl (SZ)}
 $^{1}$Astrophysics Research Centre and Observatory, \c{C}anakkale
 Onsekiz Mart University,\\ Terzio\u{g}lu Kamp\"{u}s\"{u}, TR-17020,
 \c{C}anakkale, Turkey\\
 $^{2}$Department of Physics, Faculty of Arts and Sciences,
 \c{C}anakkale Onsekiz Mart University,\\ Terzio\u{g}lu
 Kamp\"{u}s\"{u}, TR-17020,
 \c{C}anakkale, Turkey\\
 $^{3}$Astronomical Observatory, Jagiellonian University, ul. Orla
 171, 30-244 Krakow,
 Poland\\
 $^{4}$Mt. Suhora Observatory, Pedagogical University, ul.
 Podchorazych 2, 30-084 Krakow, Poland}

\begin{document}

\pagerange{\pageref{firstpage}--\pageref{lastpage}} \pubyear{2009}

\maketitle

\label{firstpage}

\begin{abstract}
This paper presents results derived from analysis of new
spectroscopic and photometric observations of the chromospherically
active binary system CF Tuc. New high-resolution spectra, taken at
the Mt. John University Observatory in 2007, were analyzed using two
methods: cross-correlation and Fourier--based disentangling. As a
result, new radial velocity curves of both components were obtained.
The resulting orbital elements of CF Tuc are:
$a_{1}{\sin}i$=$0.0254\pm0.0001$ AU,
$a_{2}{\sin}i$=$0.0228\pm0.0001$ AU, $M_{1}{\sin}i$=$0.902\pm0.005$
$M_{\odot}$, and $M_{2}{\sin}i$=$1.008\pm0.006$ $M_{\odot}$. The
cooler component of the system shows H$\alpha$ and CaII H \& K
emissions. Using simultaneous spectroscopic and photometric
observations, an  anti-correlation between the H$\alpha$ emission
and the $BV$ light curve maculation effects was found. This
behaviour indicates a close spatial association between photospheric
and chromospheric active regions. Our spectroscopic data and recent
$BV$ light curves were solved simultaneously using the
Wilson-Devinney code. A dark spot on the surface of the cooler
component was assumed to explain large asymmetries observed in the
light curves. The following absolute parameters of the components
were determined: $M_{1}$=$1.11\pm0.01$ $M_{\odot}$,
$M_{2}$=$1.23\pm0.01$ $M_{\odot}$, $R_{1}$=$1.63\pm0.02$
$R_{\odot}$, $R_{2}$=$3.60\pm0.02$ $R_{\odot}$,
$L_{1}$=$3.32\pm0.51$ $L_{\odot}$ and $L_{2}$=$3.91\pm0.84$
$L_{\odot}$. The primary component has an age of about 5  Gyr and is
approaching its Main Sequence terminal age. The distance to CF Tuc
was calculated to be $89\pm6$ pc from the dynamic parallax,
neglecting interstellar absorption, in agreement with the Hipparcos
value. The orbital period of the system was studied using the $O-C$
analysis. The $O-C$ diagram could be interpreted in terms of either
two abrupt changes or a quasi-sinusoidal form superimposed on a
downward parabola. These variations are discussed by reference to
the combined effect of mass transfer and mass loss, the Applegate
mechanism and also a light-time effect due to the existence of a
third body in the system.
\end{abstract}

\begin{keywords}
binaries: eclipsing -- stars: fundamental parameters -- technique:
spectroscopy -- technique: photometry -- stars: individual (CF Tuc)
\end{keywords}

\section{Introduction}

Recently, the study of chromospherically active binaries (hereafter
CAB) has become more popular (e.g. \cite{e} for a third
version of the catalogue of CABs; \cite{z1b} for SZ Psc;
\cite{fb} for $\lambda$ And and II Peg; \cite{s3}
for HD 6286; \cite{fa} and later series of papers). CAB
stars are usually either detached or semi-detached systems and their
components include late spectral types of F, G or K, with luminosity
classes V or IV. They show large asymmetries in their light curves,
that are usually explained in terms of cool/dark spot(s) models. The
chromospheric activity is associated with CaII H and K or/and
H$\alpha$ emission lines, ultraviolet excess, soft X-rays and radio
emission. There is a wealth of associated phenomena - including
starspots, plages, flares, non-radiatively heated outer atmospheres,
activity cycles, and deceleration of rotation rates.

The study of such active binary stars provides understanding of the
origin, evolution, and effects of magnetic fields in cool stars. In
this context, we have chosen CF Tuc as the target of the present
work. A review of this binary is presented below, new spectroscopic
observations and their reductions are described in Section~2. The
procedure used to obtain radial velocities and the orbital solution
are outlined in Section~3, while rotation velocities of the
components are examined in Section~4. Magnetic activity indicators,
H$\alpha$ and CaII H \& K emission lines, are examined in Section~5,
while simultaneous solutions of the $BV$ light and radial velocity
curves are given in Section~6. The orbital period analysis is
reviewed in Section~6, and a discussion of the new results is given
in the last section.

CF Tuc (HD 5303 = HIP 4157, $V$ = 7.60 mag) is a well known southern
hemisphere CAB. In the third edition of the CAB catalogue, recently
published by \cite{e},  the following physical properties are given:
(i) It  is a RS CVn-type SB2 system with the spectral types
G0V/IV+K4IV/V (\cite{c3e}). (ii) Photometric light curves have
remarkable asymmetries, especially visible well out of eclipses. The
magnitudes of these asymmetries are variable with time (e.g.
\cite{b4a}; \cite{a}). (iii) Radial velocities of the components
were obtained, and then spectroscopic orbital elements and absolute
parameters of the components were derived: $M_{1}$=1.06 $M_{\odot}$,
$M_{2}$=1.21 $M_{\odot}$, $R_{1}$=1.67 $R_{\odot}$, $R_{2}$=3.32
$R_{\odot}$, $T_{1}$=5980 K and $T_{2}$=4210 K (\cite{c3c};
\cite{c3a}). (iv) The orbital period is 2.797672 days, however it is
changing (\cite{t2}; \cite{a}; \cite{h5}, \cite{h6a}). (v) Orbital
inclination is close to 70$^{\circ}$, therefore eclipses are partial
(e.g. \cite{b4a}; \cite{a}). (vi) The projected rotational
velocities of the components are around 25 -- 30 km/s for the
primary, 52 - 75 km/s for the secondary (e.g. \cite{c3b}). (vii) The
spectra of the system give relatively high Li abundance
(\cite{r1a}). (viii) The system shows strong CaII H and K emission
(e.g. \cite{h2}). The H$\alpha$ line appears as nearly totally
filled-in absorption (\cite{c3d}). (ix) The system is also a known
radio and X-ray source (\cite{b4}; \cite{p2}). Some large flare
events were reported (e.g. \cite{k}). Other references can be found
in the CAB catalogue.

Particular aims of the present study are: (i) to investigate the
orbital period changes of CF Tuc with updated data, and (ii) to
derive more accurate absolute parameters of the system using our
new, high resolution spectroscopic data.

\section{Spectroscopic observations}

The spectroscopic observations of CF Tuc were made at Mt John
University Observatory (hereafter MJUO) in New Zealand in the summer
season of 2007. A 1m McLennan telescope and HERCULES (High
Efficiency and Resolution Canterbury University Large \'{E}chelle
Spectrograph) spectrograph were used. The spectrograph has 100
\'{e}chelle orders, which cover wavelength range between 380 nm and
900 nm, and is located in a vacuum tank on stable ground in a
thermally isolated room and attached to the telescope by fibre
cables. There are two different resolving powers: R=41000 and
R=70000. In the observations the former was used as being better
adjusted to the mean seeing value ($\theta$ $\sim$ ${3.5''}$) at
MJUO given by \cite{h3}.

\begin{table*}
\caption{Journal of spectroscopic observations of CF Tuc.
Signal-to-noise (S/N) ratio refers to the continuum near 5800\,\AA.}
\label{table1}
\begin{tabular}{ccccc}
\hline
No  &  Frame   &    HJD     &   S/N      & \multicolumn{1}{c}{Exp.Time} \\
    &          & (+2450000) &            & \multicolumn{1}{c}{(s)}   \\
\hline
 1 & w4350022 &  54349.98652 &  60  & 1509 \\
 2 & w4350030 &  54350.03238 &  74  & 1223 \\
 3 & w4350053 &  54350.23328 &  78  & 1322 \\
 4 & w4351012 &  54350.89945 &  85  & 1144 \\
 5 & w4351014 &  54350.91536 &  80  & 1021 \\
 6 & w4352010 &  54351.90655 &  50  & 1500 \\
 7 & w4352019 &  54351.96099 &  78  & 1500 \\
 8 & w4352022 &  54351.98189 &  85  & 1200 \\
 9 & w4353002 &  54352.86852 &  85  & 1500 \\
10 & w4354017 &  54353.93546 &  80  & 1300 \\
11 & w4354019 &  54353.95324 &  85  & 1300 \\
12 & w4354037 &  54354.10935 &  95  & 1300 \\
13 & w4354039 &  54354.12710 &  88  & 1300 \\
14 & w4356003 &  54355.85715 &  55  & 1600 \\
15 & w4356005 &  54355.87984 &  60  & 1700 \\
16 & w4356018 &  54355.99775 &  70  & 1500 \\
17 & w4356028 &  54356.07125 &  80  & 1500 \\
18 & w4356030 &  54356.09149 &  80  & 1500 \\
19 & w4363011 &  54362.85380 &  98  & 1600 \\
20 & w4363027 &  54362.93923 &  80  & 1900 \\
21 & w4363029 &  54362.96309 &  85  & 1750 \\
22 & w4377014 &  54376.89666 &  94  & 2100 \\
23 & w4377022 &  54376.96563 &  78  & 2200 \\
24 & w4378016 &  54377.94931 &  82  & 2100 \\
\noalign{\smallskip} \hline\noalign{\smallskip}\\
\end{tabular}
\end{table*}

24 spectra were obtained over 10 nights between September and
October 2007. The journal of observations is shown in Table~\ref{table1}.
The exposure time was chosen between 1100 and 1800 seconds, depending on
the weather conditions. During the observations, comparison spectra
of a thorium-argon arc lamp were taken before and after each stellar
image. A set of white lamp images was also taken as flat field
images. Two IAU radial velocity standard stars (HD 36079 and HD 693)
were observed. The bright, non-active and slowly rotating standard
star HD 36079 (G5II, $V_{r}$=-13.6 km/s) was chosen for the RV
measurements of the components of CF Tuc.  Hercules Reduction
Software Package (HRSP, ver.~3: Skuljan and Wright 2007) was used
for reductions of all observations. This procedure takes into
account other sources of apparent motion (such as the Earth's
rotations and orbital revolution), wavelength calibration, removal
of spurious pixels (such as those struck by cosmic rays), correction
for the inherent pixel-to-pixel response variation (flat-fielding),
continuum normalization, etc., and produces a target spectrum
(wavelength versus flux) as the result.

\section{Radial velocities and the orbital solution}

Measurements of radial velocities were done by two methods: The
cross-correlation technique (CCT) and the Fourier disentangling
technique (KOREL). CCT was used as the first step to estimate the
orbital parameters, as KOREL may not produce a unique solution in
some complex cases. In this study, the real orbit improvement comes
from usage of KOREL and the number of spectral orders is not so
relevant to accuracy. In general, if all the spectral orders were of
the same quality, the precision of a measurement would increase with
the square root of the number of orders. If the best spectral orders
are chosen, others will produce systematic errors, and improvement
will not be scaled as the square root of the number of orders. The
main issue is the role of noise and better line resolution and
identification in the selected orders. In the case of CF Tuc, since
CF Tuc is rather faint for the used observational instruments and
its components are close to G0 V/IV + K4 V/IV, the observed spectra
of the system include too many blended metallic absorption lines. It
is very difficult to identify real/true lines and to resolve them
into two components. All spectral regions were examined, and, as a
result, four spectral orders were selected, for which the lines of
both components could be clearly detected, and therefore suitable
for disentangling. The information about these four spectral orders
is given in Table~\ref{table2}.

\begin{table}
\begin{center}
 \caption{Spectral orders and stellar lines used in RVs measurements of CF Tuc.}
 \label{table2}
\begin{tabular}{lll}
\hline
Order No  &  Wavelength    &    Dominant      \\
          & Interval (\rm\AA)           & Spectral Lines \\
\hline 85        &  6640-6740 & SiII (6660.52 \rm\AA), SiII (6665.0
\rm\AA) \\
    &   & FeI (6677.989 \rm\AA), SiII (6717.04 \rm\AA)  \\

88        &  6430-6440 & FeI (6430.844 \rm\AA), CaI (6439.075
\rm\AA), CaI (6449.81 \rm\AA) \\
    &   & CaII (6456.87 \rm\AA), CaI
(6462.57\rm\AA)   \\
97        &  5820-5900 &   NaI D2 (5892 \rm\AA), NaI D1 (5898\rm\AA)     \\
110       &  5151-5188 &  MgI (5174.13 \rm\AA), FeI
(5168.897\rm\AA), FeII (5169.03 \rm\AA)  \\
    &   & FeII (5171.595 \rm\AA), MgI (5172.6843 \rm\AA), MgI (5174.13 \rm\AA)  \\
\hline
\end{tabular}
\end{center}
\end{table}

For the CCT, the FXCOR task in the radial velocity package of IRAF
(\cite{t3}; \cite{p1}) was used. FXCOR calculates the velocity
Doppler shift between two spectra (of the variable and comparison
stars) by fitting the correlation with a user-selected function. In
the present study, the Gaussian function was adopted as the
best-fitting one. The spectra of HD 36079 were used as a template
for deriving RVs of the components. In order to obtain orbital
parameters from the radial velocity data derived from the CCT, the
ELEMDR77 program, developed by T. Pribulla (2008, private
communication), was used.

\begin{figure*}
\begin{center}
\begin{tabular}{cc}
      \resizebox{80mm}{!}{\includegraphics{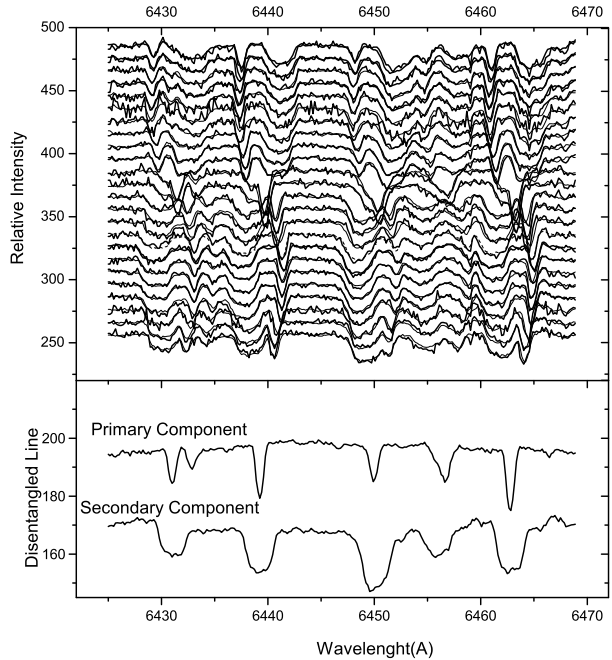}} \\ \\
\end{tabular}
\caption{KOREL solution of echelle order 88 as a sample. Upper panel
shows the KOREL fits while the lower panel displays the disentangled
spectra of each component.} \label{fig1}
\end{center}
\end{figure*}

For the Fourier disentangling technique, the KOREL code (developed
by \cite{h1a}, \cite{h1b}) was used. In the first step with this
procedure, KOREL requires input parameters within realistic bounds.
The initial values of parameters were taken from the solution of RV
curves obtained with the CCT method. Four spectral orders, given in
Table \ref{table2}, were analyzed simultaneously. After several
iterations, the KOREL code gave a value close to 0 for the
eccentricity $e$ within its uncertainties. For this reason, we
assumed a circular orbit for the system. Additionally, during the
fitting, the orbital period $P_{orb}$ of the system was fixed to be
2.7975004 days (see Section~5). The velocity amplitudes $K_1$ and
$K_2$ of the components and the conjunction time $T_0$ were the
adjusted parameters. The best fitting orbital elements are given in
Table~\ref{table3}, and the best fits to the composite spectra and
disentangled spectra are shown in Fig. \ref{fig1} for the echelle
order 88 as a sample.

\begin{table}
\caption{Spectroscopic orbital parameters of CF Tuc} \label{table3}
\begin{tabular}{lc}
\hline
\textbf{Parameter} & \textbf{Value}\\[2pt]
\hline
$P_{orb}$ (days) & 2.7975004 (fixed) \\
$T_{o}$ (HJD)       & 54327.0583${\pm}$0.0012 \\
$V_{\gamma}$(km/s)   & 9.58${\pm}$0.14  \\
$q$            & 1.117${\pm}$0.009 \\
$K_{1}$(km/s)    & 98.92${\pm}$0.24 \\
$K_{2}$(km/s)    & 88.55${\pm}$0.24 \\
$a_{1}{\sin}i$(AU) & 0.0254${\pm}$0.0001 \\
$a_{2}{\sin}i$(AU) & 0.0228${\pm}$0.0001 \\
$M_{1}{\sin}i$(${M_{\odot}}$) & 0.902${\pm}$0.005 \\
$M_{2}{\sin}i$(${M_{\odot}}$) & 1.008${\pm}$0.006 \\
\hline
\end{tabular}
\end{table}

\begin{table*}
\begin{center}
\caption{RV measurements, with O--C values from theoretical fit, of
components of CF Tuc.} \label{table4}
\begin{tabular}{ccrcrr}
        &           &            &              &             &            \\
\hline
Time    &   Phase   &  $RV_{1}$  &   $O-C_{1}$  &   $RV_{2}$  &   $O-C_{2}$ \\
HJD     &   $\phi$  &(km s$^{-1}$) & (km s$^{-1}$)  & (km s$^{-1}$) &   (km s$^{-1}$) \\
\hline
2454377.9493 & 0.192 & -82.4 &  1.0  & 91.8 &  -0.5   \\
2454349.9865 & 0.196 & -84.6 & -0.4 &  92.3 &  -0.8    \\
2454350.0324 & 0.212 & -87.2 & -0.2 &  96.2 &   0.8     \\
2454352.8686 & 0.226 & -87.9 &  0.7 &  96.8 &   0.0  \\
2454350.2333 & 0.284 & -87.7 & -0.7 &  95.5 &   0.1  \\
2454355.8572 & 0.295 & -85.3 &  0.0 &  94.4 &   0.6  \\
2454355.8798 & 0.303 & -84.4 & -0.7 &  92.6 &   0.1  \\
2454355.9978 & 0.345 & -72.5 & -0.7 &  87.0 &   4.7  \\
2454356.0713 & 0.371 & -62.1 & -0.6 &  73.7 &   0.3  \\
2454356.0915 & 0.378 & -58.9 & -0.5 &  70.8 &   0.1  \\
2454350.8994 & 0.522 &  24.9 &  0.5 &     -  &   -    \\
2454350.9154 & 0.528 &  26.7 & -1.3 &     -  &   -   \\
2454353.9355 & 0.608 &  71.1 & -1.7 & -45.4 &   0.9  \\
2454353.9532 & 0.614 &  75.7 &  0.1 & -48.2 &   0.6  \\
2454354.1093 & 0.670 &  95.8 & -1.3 & -67.5 &   0.6  \\
2454354.1271 & 0.676 &  96.0 & -2.8 & -71.8 &  -2.2  \\
2454362.8538 & 0.796 & 104.3 & -0.1 & -75.1 &   0.3  \\
2454376.8967 & 0.815 & 100.2 &  0.0 & -70.4 &   1.4 \\
2454362.9392 & 0.826 &  97.3 &  0.0 & -68.8 &   0.3 \\
2454362.9631 & 0.835 &  94.8 &  0.3 & -66.6 &   0.1 \\
2454376.9656 & 0.840 &  93.0 &  0.2 & -65.0 &   0.2 \\
2454351.9065 & 0.882 &  76.3 &  0.4 & -50.0 &  -0.1 \\
2454351.961  & 0.902 &  66.9 &  0.9 & -41.7 &  -0.5 \\
2454351.9819 & 0.909 &  63.1 &  0.7 & -38.2 &  -0.4 \\
\hline \\
\end{tabular}
\end{center}
\end{table*}

The KOREL code can not derive the systemic velocity of the binary
star, however, KOREL retains the systemic velocity in the
disentangled spectra of the components. Therefore, the systemic
velocity $V_{\gamma}$ was adopted by averaging systemic velocities
obtained from the CCT method and the ELEMDR77 program. The adopted
systemic velocity was added to the RVs measured by KOREL to yield
the final RVs of the components, which are given in
Table~\ref{table4}. In this table, phase values of observed time of
RVs in the second column were calculated using the linear ephemeris
given in Eq.~1. $O$-$C$ values in the fourth and sixth columns
represent the residuals between observed and theoretical RVs
obtained from simultaneous solution of the $BV$ light and RV curves,
described in Section~5.

\section{Rotational velocities}

The program PROF (\cite{b4a}), which follows an ILOT
type curve-fitting procedure, was used to determine rotational
velocities. PROF convolves Gaussian and rotational broadenings, as
discussed by \cite{b4a}, and computes the line
profile as a function, basically, of the following parameters: the
continuum intensity $I_{c}$, the relative depth $I_{d}$ at mean
wavelength $\lambda_{m}$, the rotational broadening parameter $r$,
Gaussian broadening parameter $s$ of a given line, and the limb
darkening coefficient $u$. A similar procedure was followed by
\cite{o}, \cite{o2}  and \cite{b4b}.

\begin{figure}
\begin{center}
      \resizebox{90mm}{!}{\includegraphics{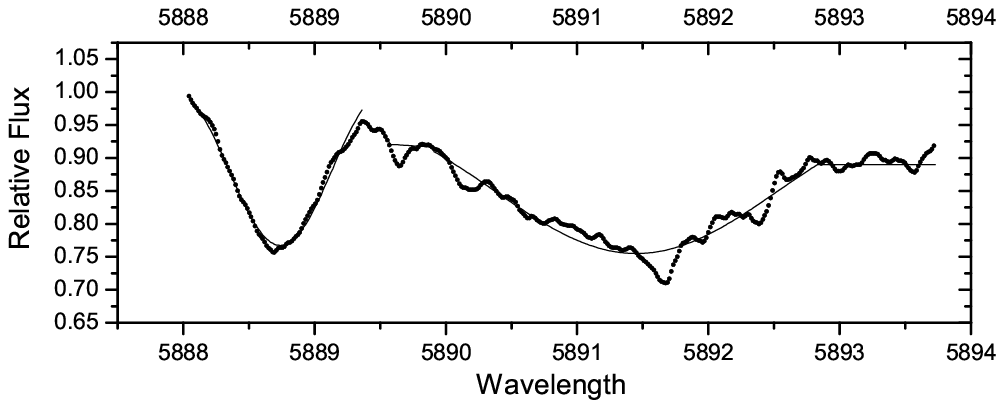}} \\
      \resizebox{90mm}{!}{\includegraphics{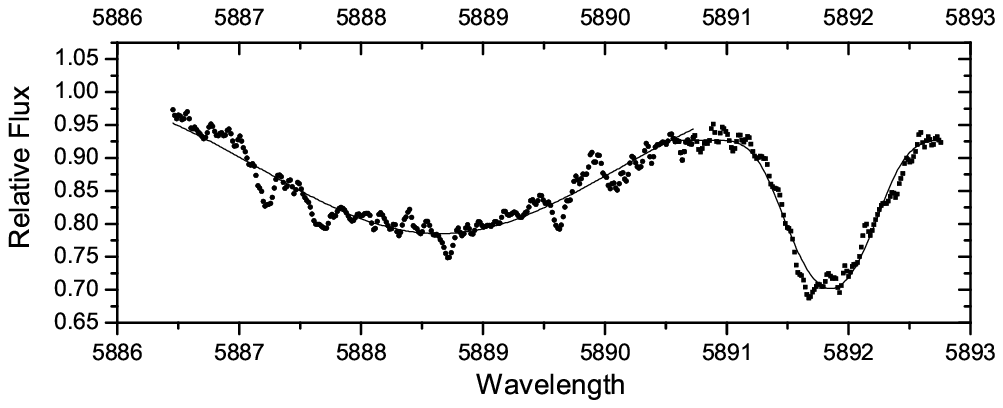}} \\
\caption{Results of profile fitting to Na D2 5892 lines at phase
0.371 (upper panel) and phase 0.826 (lower panel).} \label{fig2}
\end{center}
\end{figure}

We considered the Na D2 line profiles for CF Tuc in the Hercules
spectral order 97 and fitted the selected line profiles at various
orbital phases using PROF. Typical results of the profile fitting at
phases of 0.371 and 0.826 are shown  in Fig.~\ref{fig2} and given in
Table~\ref{table5}.

\begin{table}
   \caption{Profile fitting parameters at two different orbital phases for the Na D2 5892 feature.} \label{table5}
\begin{tabular}{llllll}
\hline
& \multicolumn{2}{c}{phase 0.371} & \multicolumn{2}{c}{phase 0.826} \\
Parameter               & Primary               & Secondary & Primary & Secondary \\
\hline
$I_{c}$                 & 1.024$\pm$0.015       & 0.990$\pm$0.010       & 0.927$\pm$0.012       & 1.031$\pm$0.008 \\
$I_{d}$                 & 0.216$\pm$0.018       & 0.188$\pm$0.010       & 0.192$\pm$0.019       & 0.175$\pm$0.008 \\
$\lambda_{m}$(\rm\AA)   & 5888.740$\pm$0.043    & 5891.452$\pm$0.094    & 5891.852$\pm$0.043    & 5888.641$\pm$0.102 \\
$r$(\rm\AA)             & 0.498$\pm$0.053       & 1.142$\pm$0.108       & 0.528$\pm$0.055       & 1.216$\pm$0.088 \\
$s$(\rm\AA)             & 0.469$\pm$0.107       & 0.745$\pm$0.074       & 0.300$\pm$0.060       & 1.053$\pm$0.055 \\
\hline
$r$(km/s)               & 25$\pm3$              & 58$\pm6$              & 27$\pm3$              & 62$\pm5$ \\
$s$(km/s)               & 24$\pm$6              & 38$\pm$4              & 15$\pm$3              & 54$\pm$3 \\
\hline
$\Delta l$              & 0.01                  & 0.01                  & 0.01                  & 0.01 \\
$\chi^{2}/\nu$          & 1.034                 & 1.040                 & 1.024                 & 1.048 \\
\hline \\
\end{tabular}
\end{table}

According to the value of $r$ (the rotational broadening parameter)
in Table~\ref{table5}, the projected rotational velocities of the primary and
secondary components are 26$\pm3$ and 60$\pm5$ kms$^{-1}$,
respectively. Using absolute parameters of components
from Table~\ref{table8} and $v_{rot}$sin$i$=2$\pi R$sin$i$/$P_{rot}$ (assuming
synchronous rotation, $P_{rot}$=$P_{orb}$, and $i_{rot}$=$i_{orb}$),
theoretical rotational velocities were be found as 28$\pm1$ and
62$\pm2$ kms$^{-1}$ for the primary and secondary components,
respectively. Therefore, we found that within the error limits,
synchronous rotation for both components can be reliably accepted.

The Gaussian broadening parameter $s$ (presented in
Table~\ref{table5}) ranges from 24 to 15 km/s for the primary
component and from 38 to 54 km/s for the secondary component at
phases of 0.371 and 0.826, respectively. This parameter generally
relates thermal broadening and other broadening factors (micro
and/or macro turbulence, etc.) rather then rotational broadening. If
the temperature of 4300 K is taken, where the Na D2 lines are formed
in a subgiant atmosphere, thermal velocities are found to be $\sim$2
km/s. Therefore, for this value, the thermal broadening is not
significant, and the broadening parameter must include some other
motions, possibly related to turbulence effects or magnetic
activity. Since our spectroscopic observations and Innis's CCD $BV$
observations were made almost simultaneously, we can compare these
spectroscopic results with the spot model derived from the $BV$
light curve analyses. For instance, when the maculation is close at
phase 0.821, the $s$ parameter is greatest and falls to about 38
km/s half a cycle later (at phase 0.371). Therefore, we can deduce
that the larger value of $s$ of the secondary component could be
associated with the relatively stronger effects of surface activity
of this star.

\section{Magnetic activity indicators}

The H$\alpha$ and CaII H \& K emission lines are very important
indicators of magnetic activity - in other words, chromospheric
activity. Generally, the more active stars show these emission lines
always above the continuum (e.g. UX Ari, II Peg, AR Psc, V711 Tau
and XX Tri). Except for  being sensitive to the chromospheric activity,
these emission lines are also a good diagnostic of inter-components
matter in the form of gas streams, transient or classical accretion
disks and rings in mass-transferring binaries i.e. the
Algol type, in which the cooler star fills its Roche lobe and
transfers mass to the hot companion (e.g. \cite{r1c}).

\begin{figure}
\begin{center}
      \resizebox{80mm}{!}{\includegraphics{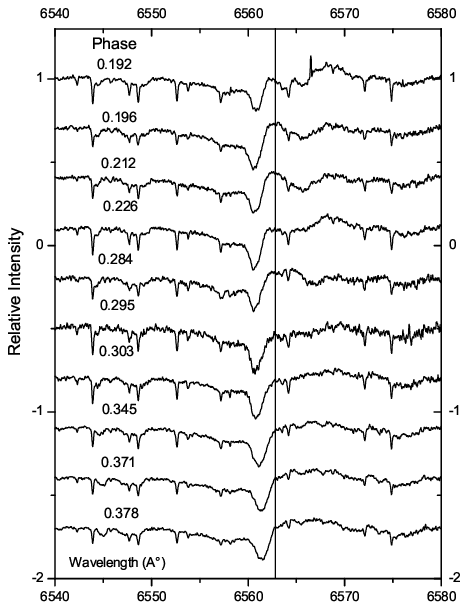}}
      \resizebox{80mm}{!}{\includegraphics{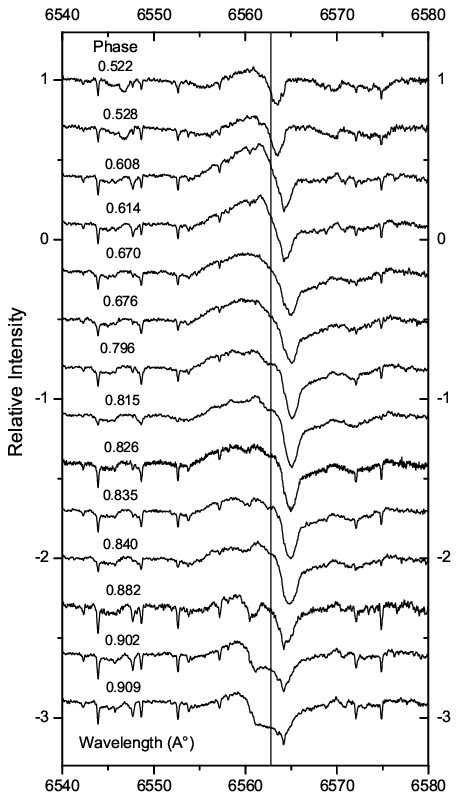}} \\
\caption{Observed H$\alpha$ line profiles for CF Tuc in 2007 at
first half (left panel) and second half (right panel) of the orbital
period. The vertical lines represent the laboratory wavelength
($\lambda_{0}$=6562.82 {\AA}) of the H$\alpha$ line.} \label{fig3}
\end{center}
\end{figure}

\begin{figure}
\begin{center}
      \resizebox{80mm}{!}{\includegraphics{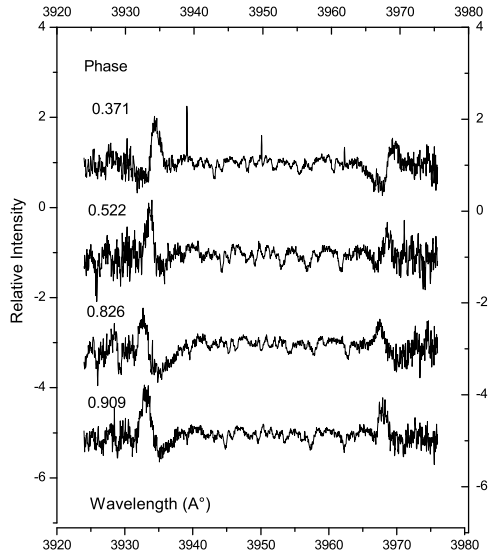}}
\caption{Sample of observed CaII H \& K spectra for CF Tuc at
different orbital phases.} \label{fig4}
\end{center}
\end{figure}

\begin{figure}
\begin{center}
      \resizebox{80mm}{!}{\includegraphics{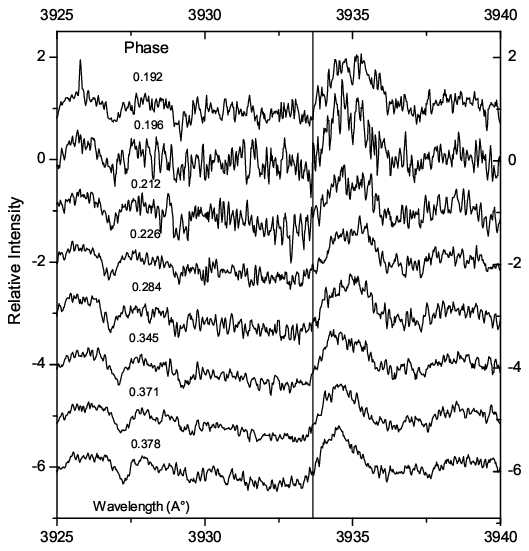}}
      \resizebox{80mm}{!}{\includegraphics{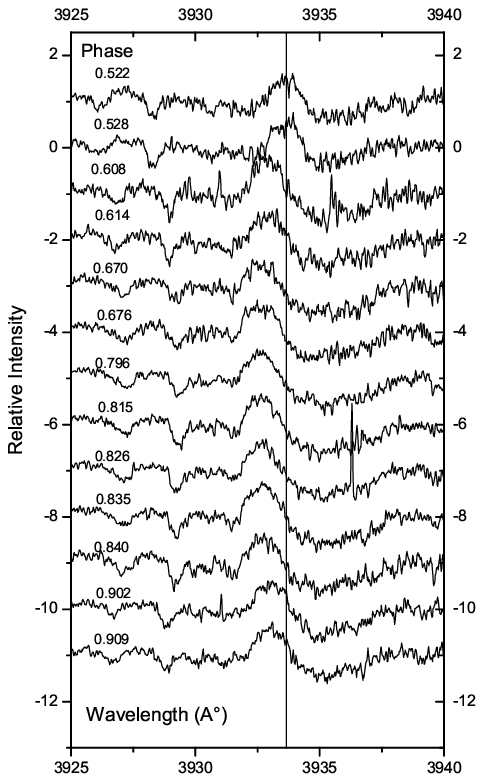}} \\
\caption{Observed CaII K line profiles for CF Tuc in 2007 at first
half (left panel) and second half (right panel) of the orbital
period. The vertical lines represent the laboratory wavelength
($\lambda_{0}$=3933.66 {\AA}) of the CaII K line.} \label{fig5}
\end{center}
\end{figure}

\begin{figure}
\begin{center}
      \resizebox{100mm}{!}{\includegraphics{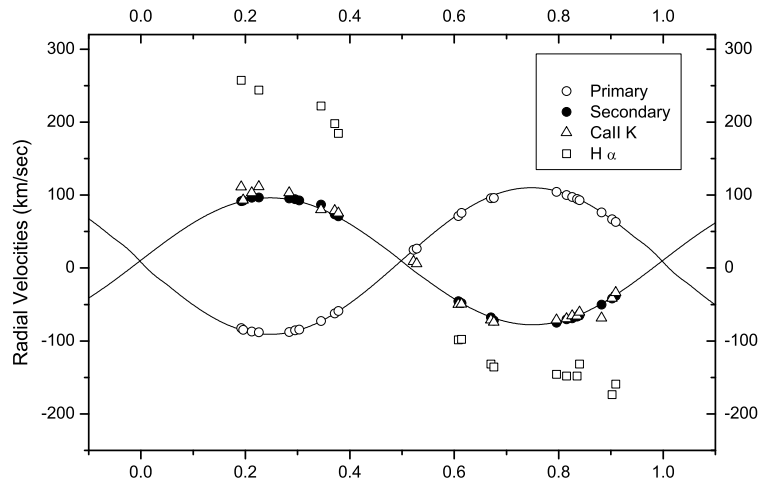}} \\
\caption{Variations of RVs of emission features observed in
H$\alpha$ and CaII K. The continuous lines are the RV solutions for
the primary (hotter) and secondary (cooler) components, according to
parameters given in Table~6.} \label{fig6}
\end{center}
\end{figure}

The H$\alpha$ and CaII H \& K observed line profiles of CF Tuc are
shown in Figs.~\ref{fig3}, \ref{fig4} and \ref{fig5}.
In these spectra, the H$\alpha$ profiles
show absorption and emission features, while the CaII H \& K
are in  emission at all orbital phases. All
absorption and emission features were red and blue-shifted depending
on the orbital phases. Since the cooler component of the system is
an active star, it is assumed that these emission features are
related to this component. To confirm this, we calculated the RV
values of the H$\alpha$ and CaII K emission features and plotted
them in Fig.~\ref{fig6} with RVs of both components of the binary system. We
took into account only the K line (3933.66 {\AA}) here, because of
likely contamination from H$\epsilon$ centered only 1.5 {\AA} away
from the H line (3968.47 {\AA}). As it can be seen in Fig.~\ref{fig6}, although
not strictly sinusoidal, the RVs of the H$\alpha$ emission feature
follows the orbital motion of the cooler component but with a larger
amplitude of about 200 kms$^{-1}$. This rather larger amplitude
indicates that the H$\alpha$ emission feature could originate in a
gas cloud in the form of a chromospheric prominence from the cooler
component. The RVs of the CaII K emission line closely follows that
of the cooler component (see Fig.~\ref{fig6}), indicating  its origin to be
in the chromospheric layers of that star.

\begin{figure}
\begin{center}
      \resizebox{80mm}{!}{\includegraphics{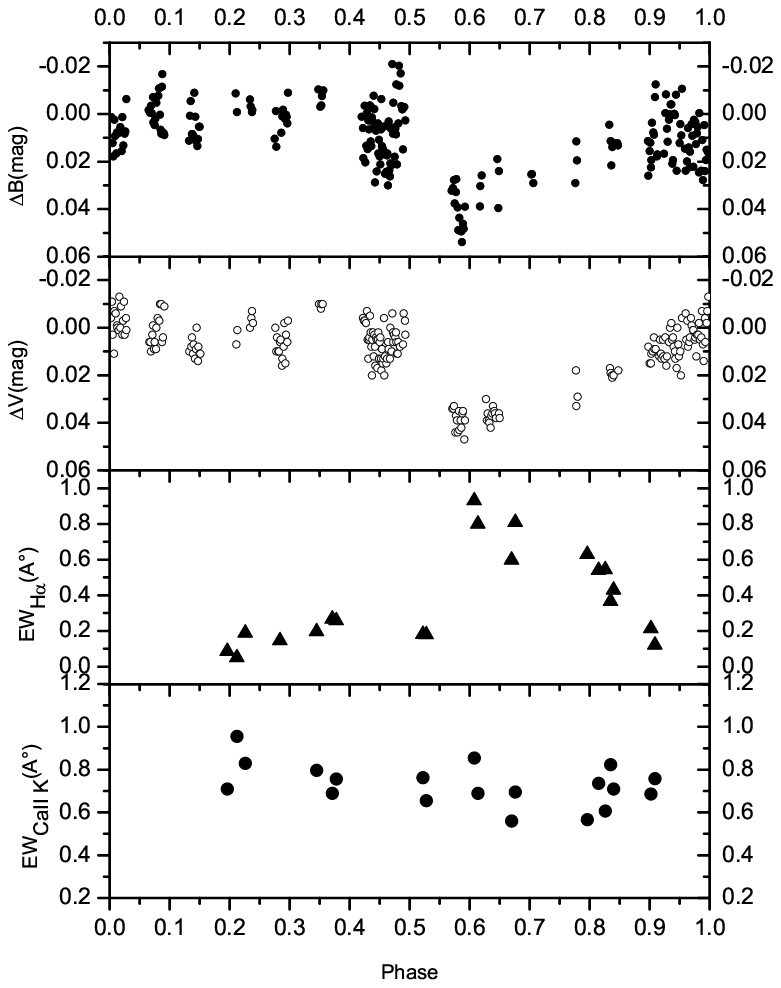}} \\
\caption{Maculation effects in $BV$ magnitudes (upper two panels),
and observed $EW$s of H$\alpha$ and CaII K emission features (lower
two panels) as a function of the orbital phase for CF Tuc.}
\label{fig7}
\end{center}
\end{figure}

If we assume that the H$\alpha$ emitting region is at rest in the
rotating reference frame of the system and that it lies between the
barycentre of the system and the cooler component, we can estimate
its location following \cite{Ma}, who modelled inter-components
matter using H$\alpha$ lines for HR 7428. The HR 7428 star is an RS
CVn-type binary, composed of a bright K giant and an A-type Main
Sequence dwarf. It is a well detached system like our target star,
CF Tuc. In this model, the projected distance of the emitting region
can be found from $a_{er}$=$a_{c}$($K_{er}$/$K_{c}$), where $a_{c}$
is the semi-major axis of the cooler star orbit, and $K_{c}$ and
$K_{er}$ are the semi-amplitudes of the RV variations for the cooler
star and emitting region, respectively. Since our orbital solution
gives $K_{c}$=89 kms$^{-1}$ and $a_{c}$=3.6$\times$10$^{6}$ km, we
find $a_{er}$=8.2$\times$10$^{7}$ km - in other words, the H$\alpha$
emitting region should be located at a distance of about 2 $R_{c}$
from the surface of the cooler component.  We estimate almost the
same value for the location of the emitting region using $v_{er,
rot}$sin$i$=2$\pi a_{er}$ sin$i$/$P_{rot}$ (with the assumption of
synchronous rotation).

The simultaneous photometric and spectroscopic observations of CF
Tuc offer the possibility of studying the photospheric
(spots) and chromospheric (plages and/or prominences) active regions
of the cooler component. With this aim, we measured the equivalent
widths ($EW$s) of emission features observed in the H$\alpha$ and
CaII K spectra by means of multiple Gaussian fits, and plotted them
versus orbital phase in Fig.~\ref{fig7}. In Section 6, we present a large
cool photospheric spot on the cooler component to explain observed
$BV$ light curve asymmetries. To concentrate on effects due to the spot
only, we subtracted the eclipse and proximity effects from the
observed data using the unspotted (immaculate) light curve
parameters given in Table~\ref{table6} and plotted these points (which show the
maculation/distortion wave) versus the orbital phase in the upper
panels of Fig.~\ref{fig7}.

The maximum $EW_{H\alpha}$ value of about 1 {\AA} is reached at
phases 0.6 -- 0.7, where the spot effect is dominant ($BV$
distortion of lights minima); the minimum value of
$EW_{H\alpha}\cong$0.2 {\AA} is observed between phases 0.0 and 0.5,
where the spot could not be seen in the light curves. The CaII K
equivalent width does not show any clear orbital modulation
(rotational modulation under the synchronous rotation). However,
there is a possible anti-correlation between $BV$ light curve
asymmetries and H$\alpha$ emission, which is apparent with an almost
similar shape of the curves. This behaviour denotes a close spatial
association of photospheric and chromospheric active regions. Such
anti-correlations between photospheric and chromospheric diagnostics
are found in some active stars and have been examined by many
authors (e.g. \cite{fb1}, \cite{fb2}; \cite{b3a}).

\section{Simultaneous Solution of the $BV$ Light and Radial Velocity Curves}

\begin{figure}
\begin{center}
      \resizebox{100mm}{!}{\includegraphics{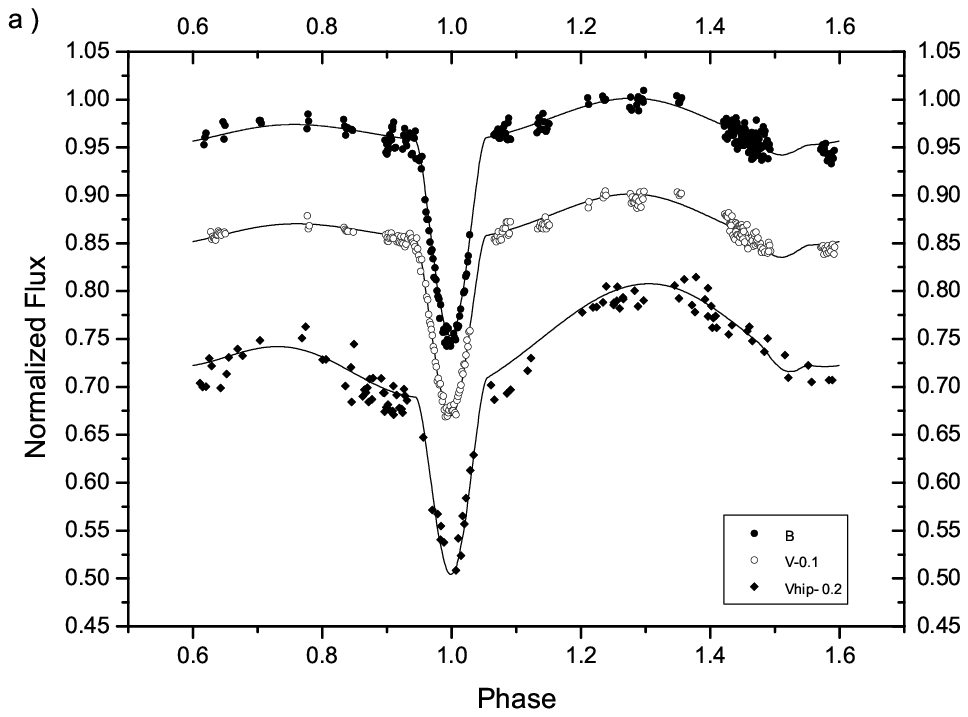}} \\
      \resizebox{50mm}{!}{\includegraphics{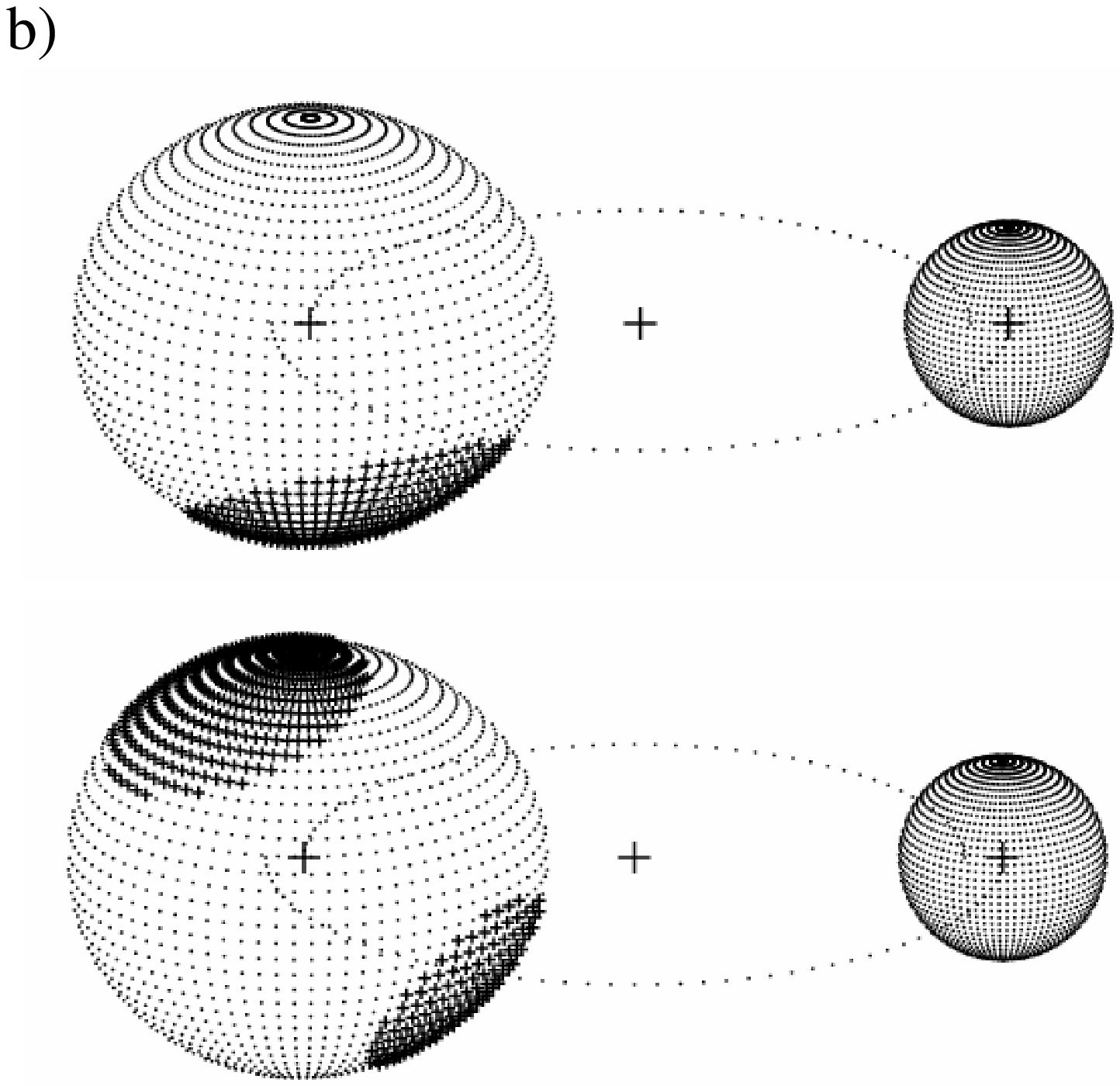}} \\
      \resizebox{70mm}{!}{\includegraphics{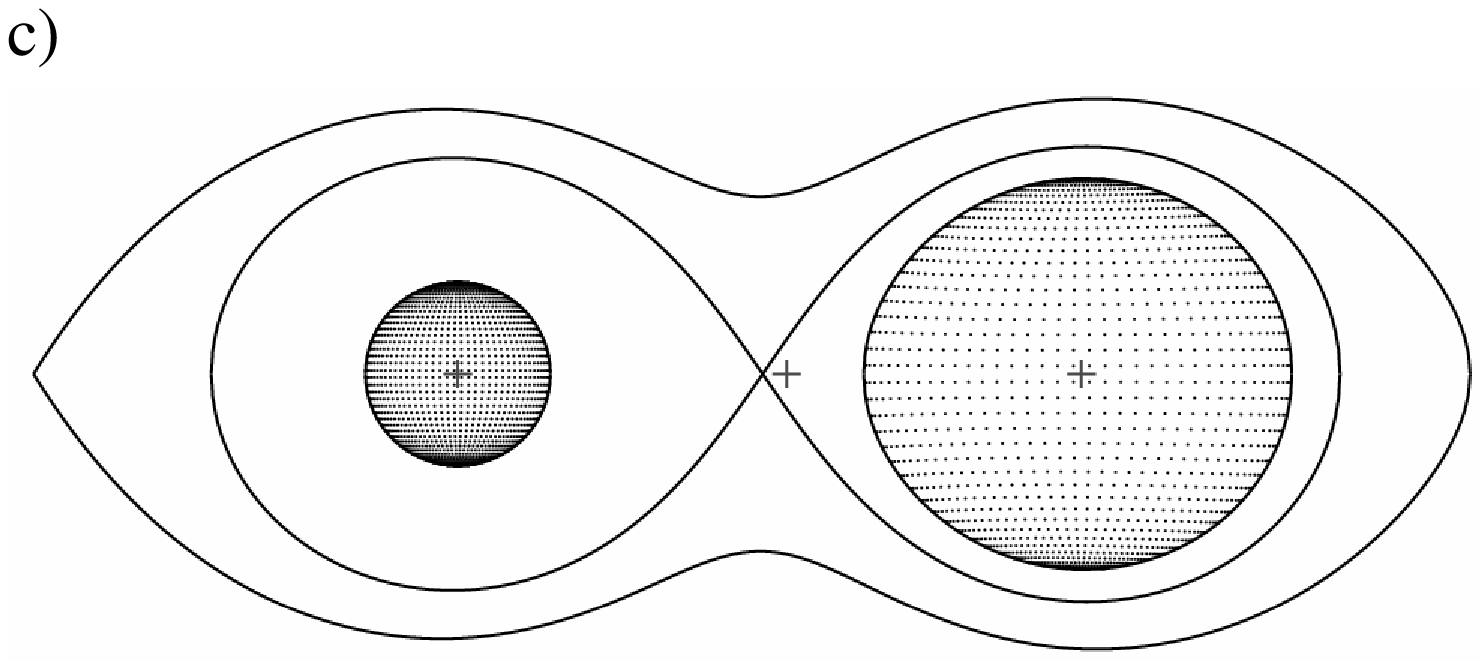}} \\
\caption{$a)$ Best theoretical fits to the $BV$ and Hipparcos light
curves, $b)$ one spot and two spots on the 3D model of the cooler
component for the $BV$ and Hipparcos light curves, respectively, and
$c)$ Roche geometry of the system.} \label{fig8}
\end{center}
\end{figure}

We analyzed the $BV$ light curves from Innis, J. L. (2008, private
communication), Hipparcos light curve (ESA 1997) and radial velocity
curves from this study using the Wilson-Devinney code (WD), version
1996 (\cite{wd71}). The $BV$ light curves from Innis and our new
radial velocity curves were solved simultaneously. Innis et al.
observed CF Tuc in $BV$ filters at Brightwater Observatory in the
summer season of 2007. They used a short-focus, 70mm telescope and a
cooled SBIG ST7E  CCD camera, which gives a field of view of near
0.8 arcdegree$\times$0.55 arcdegree. A detailed description of the
observatory and techniques is given in the paper by \cite{h6b}. They
observed HD 5210 and HD 4644 as comparison and check stars,
respectively. We calculated external uncertainties for all
comparison minus check magnitudes and found them to be 31 and 13
mmag in $B$ and $V$ filters, respectively. For this, we used the
standard deviation of the differential light variations of the
comparison relative to the check star collected during the same
night. A similar procedure was followed by \cite{e2c} and also
\cite{s2} to examine the quality of long-term multicolour
photometric data of several active stars. The observational data
were not transformed into the standard $BV$ system. It is worth noting
that our spectroscopic observations and their photometric
observations were made almost simultaneously. In order to calculate
the phases of the CCD $BV$ light observations of CF Tuc, the light
elements of the system were derived by using the photoelectric
primary minima times with $E>$2000 cycles (see Fig.~\ref{fig9}b) as;

\begin{equation}
HJD~(Min~I) = 2452452.7326(68) + 2^{\rm d}.7975004(9) \times E.
\end{equation}
with the weighted least squares method. In the light curves which
were formed using these phases, the primary minimum coincides with
the phase 0.0 (see upper panel of Fig.~\ref{fig8}).

In the WD method, some parameters could be fixed according to
theoretical models. In the light curve modelling, the temperature of
the primary component was fixed at 6100 K, following Anders et
al.(1999) and Budding and McLaughlin (1987). The root square limb
darkening law was adopted, and the darkening coefficients were taken
from \cite{d1} and \cite{c1} The bolometric gravity-darkening
coefficients of the components were set to 0.32 for convective
envelopes, following \cite{la}; also, the bolometric albedos were
fixed to 0.5 for convective envelopes, following \cite{r3}.
According to analysis of the rotational velocities (see Section~4),
the components rotate synchronously. Therefore, the rotation
parameters were assumed as F$_{h}$=F$_{c}$=1. From the spectroscopic
orbital solution described in Section~3, the circular orbit ($e$=0)
was adopted.

\begin{table*}
 \caption{Final solutions of light and radial velocity curves of CAB
star CF Tuc} \label{table6}
\begin{tabular}{lll}
\hline
Parameter & $BV+2RV$ & Hp \\
\hline
$a(R_{\odot})$      & 11.08$\pm$0.02 & -- \\
$Phase$ $shift$     & -0.0017$\pm$0.0002 & -0.0003$\pm$0.0006 \\
$V_{\gamma}$ (km/s) & 9.6$\pm$0.4 & -- \\
$i$ (deg)           & 69.91$\pm$0.09    & 69.91  \\
$T_1$ (K)           & 6100              & 6100 \\
$T_2$ (K)           & 4286$\pm$19       & 4286 \\
${\Omega}_{1}$      & 7.907$\pm$0.086   & 7.907 \\
${\Omega}_{2}$      & 4.452$\pm$0.011   & 4.452 \\
$q_{corr}=m_2/m_1$  & 1.115$\pm$0.003       & 1.115  \\
$l_1/l_{12}~(B)$    & 0.616$\pm$0.008   & -- \\
$l_1/l_{12}~(V)$    & 0.557$\pm$0.008   & 0.567$\pm$0.001 \\
$r_1(mean)$         & 0.148$\pm$0.001   & 0.148 \\
$r_2(mean)$         & 0.325$\pm$0.001   & 0.325 \\
\hline
\it Spot~parameters & \\
\hline
Spot1 co-latitude (deg)   & 155$\pm$5        & 121$\pm$5\\
Spot1 longitude (deg)     & 303$\pm$4        & 329$\pm$3 \\
Spot1 radius (deg)        & 47$\pm$2         & 31$\pm$4 \\
Spot1 T$_{spot}$/T$_{star}$ & 0.746$\pm$0.038 & 0.731$\pm$0.026 \\
Spot2 co-latitude (deg)   & --              & 24$\pm$3  \\
Spot2 longitude (deg)     & --              & 206$\pm$3 \\
Spot2 radius (deg)        & --              & 38$\pm$2 \\
Spot2 T$_{spot}$/T$_{star}$ & --            & 0.785$\pm$0.028 \\
$\Sigma W (O-C)^{2}$ & 0.03177 & 0.01192 \\
\hline
\end{tabular}
\end{table*}

The adjusted parameters in our computations are: the semi-major axis
of orbit ($a$), phase shift, systemic velocity of the binary
($V_{\gamma}$), orbital inclination ($i$), surface temperature of
the secondary component ($T_{2}$), non-dimensional surface
potentials of both components (${\Omega}_{1}$ and ${\Omega}_{2}$),
and the fractional monochromatic luminosity of the primary component
(l$_{1}$/(l$_{1}$+l$_{2}$)). The initial values of $q$, $A$ and
$V_\gamma$ were taken from the radial velocity solution (see
Section~3). Due to the probability of the existence of a third body,
resulting from the orbital period analysis of the system (see
Section~7), a third light contribution ($l_{3}$) was also considered
as a free parameter. However, we soon found its contribution to be
negligible. The binary CF Tuc, as mentioned above, is a typical RS
CVn type eclipsing binary and its light curves show distortion wave
like other systems in this group. In the $BV$ light curves, the
distortion wave appears to be an asymmetry between the light levels
of the maxima (see Fig.~\ref{fig8}a). Therefore, we had to consider a
cool spot on the secondary and allowed the spot parameters to
be adjusted.

In order to get good starting parameters for the WD code, mean
points were calculated from all individual observations. This
provided standard deviations for $B$ and $V$ filters; the highest
values were 15 mmag (the clump near phase 0.9) in $B$, and 11 mmag
(near phase 0.1) in $V$. These mean points (69 in $B$, 64 in $V$)
were used with a Monte Carlo search to find a fit to the $BV$ light
curves with a mass ratio fixed at the value derived
spectroscopically (see Table~\ref{table3}). The resulting parameters
were next set as starting values for simultaneous solution of the
$BV$ light and RV curves. Simultaneous $BV$ light and RV curves
solution was done for all individual points, assigning the same
weight and different errors for the $B$ and $V$ filters as resulted
from the calculation of mean points. Errors for the RV curves were
assigned the same as their quality was comparable. The NOISE control
parameter was set to 1.

Simultaneous convergent solutions of the \textit{BV} light and RV
curves were obtained  by iterations, until the corrections of the
parameters became smaller than their corresponding errors. The
results of the final solution are given in Table~\ref{table6}. The
comparison between observed and computed light curves is shown in
Fig.~\ref{fig8}a, while that of RV curves is presented in Fig.~6.
The three-dimensional model demonstrating the presence of a large
dark spot on the surface of the cooler component and the Roche
geometry of the system (making use of the Binary Maker program,
ver.3.0, \cite{b12}) are also shown in Fig.~\ref{fig8}b,c.

We made additional attempts to check the stability of our solution.
This was done by assigning larger errors for the photometric light
curves (especially for the $B$ light curve). Additionally, only RV
curves (adjusting only the parameters relevant to the orbit) were
solved using the WD code. It was found that the simultaneous
solution of $BV$ and RV curves resulted in the mass ratio
q=1.115$\pm$0.003, while only the RV solution gave
q=1.113$\pm$0.004. Also, the solutions with assigned larger errors
for $B$ data gave (within uncertainties) a similar value of the mass
ratio.

The Hipparcos light curve of CF Tuc is available from the Hipparcos
web page and contains 121 points with an average observational error
of 11 mmag. Before starting analysis, Hipparcos observations of the
system were transformed to Johnson $V$ magnitudes using $Hp-V$ =
0.22$\times$($B-V$) calibration given by \cite{r2}.

The following ephemeris was used to phase the Hipparcos photometric
data:

\begin{equation}
HJD ~ (Min~I) = 2448502.560 + 2^{\rm d}.79765\times E.
\end{equation}

The data was weighted by using the equation $w_i=1/\sigma_i^2$,
where $\sigma_i$ is the individual standard error of the data given
in the Hipparcos catalogue. About ten photometric points were
discarded due to their relatively large errors. We used the mean
maxima levels at phase 0.25 for the flux normalization of both the
Hipparcos and 2007 $BV$ filters data.

During the iterations, only spot parameters, phase shift, and
luminosity of the primary component were treated as free parameters;
others were adopted from the simultaneous solution of $BV$ and
radial velocity curves. As can be seen in Fig.~\ref{fig8}a, the
Hipparcos light curve shows two large asymmetries, one at about
phase 0.70 and another at the primary minimum. Therefore, the
possibility of two dark spots on the secondary component was
considered. The final results are given in Table~\ref{table6} and
displayed in Fig.~\ref{fig8}a,b.

\section{Orbital period analysis}

In order to investigate the orbital period variation of the system,
we gathered 33 minima times available from the lists compiled by
Kreiner, J. M. (2008, private communication) and \cite{a} and we
added one minimum time, which was calculated from the 2007 $BV$
light curves observed by Innis et al. (2008, private communication).
As a first step, $O-C$ values were calculated using the following
light elements, given by \cite{a}:

\begin{equation}
HJD~(Min~I) = 2444219.270 + 2^{\rm d}.797715 \times E.
\end{equation}

\begin{figure}
\begin{center}
\begin{tabular}{cc}
      \resizebox{80mm}{!}{\includegraphics{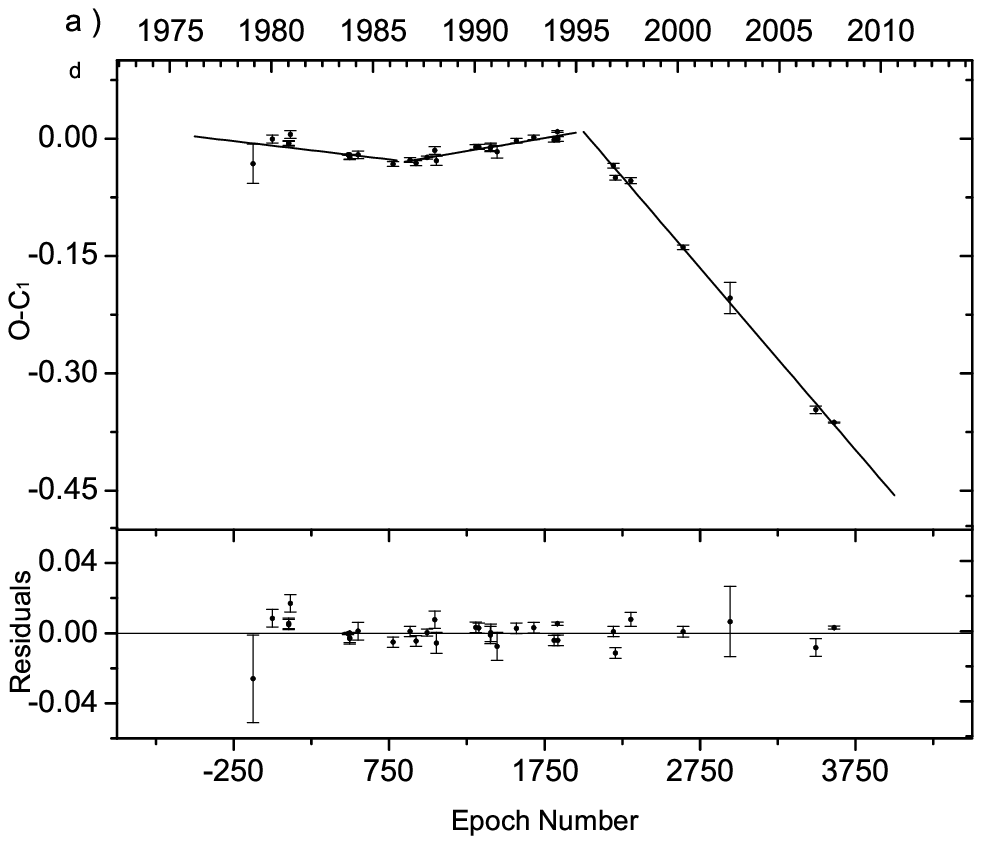}} &
      \resizebox{80mm}{!}{\includegraphics{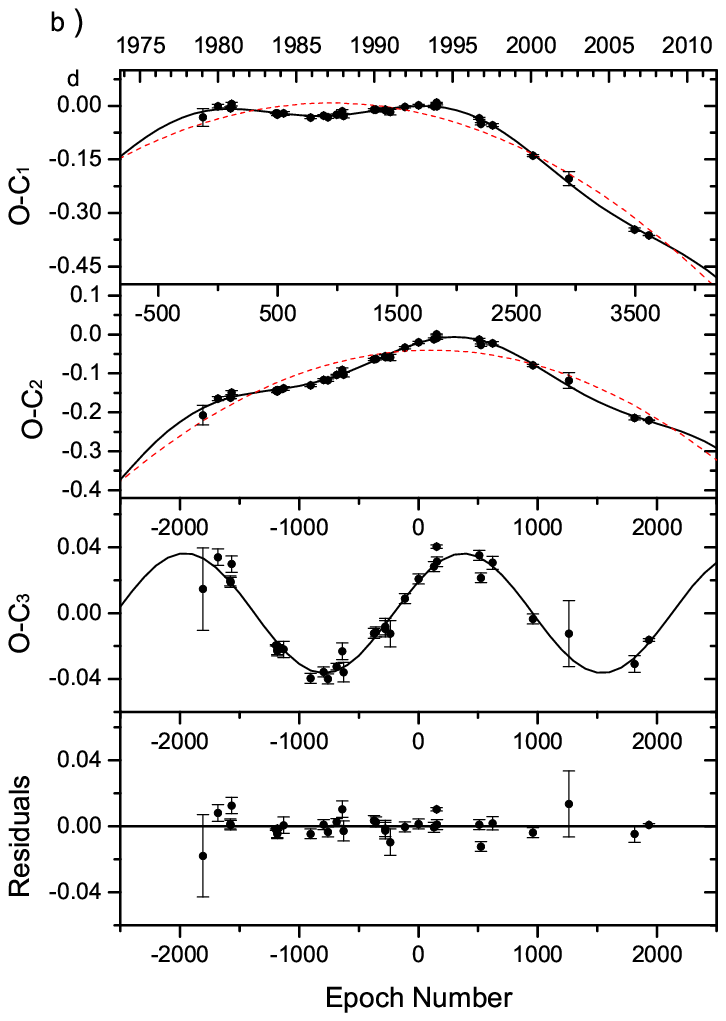}} \\
      \resizebox{80mm}{!}{\includegraphics{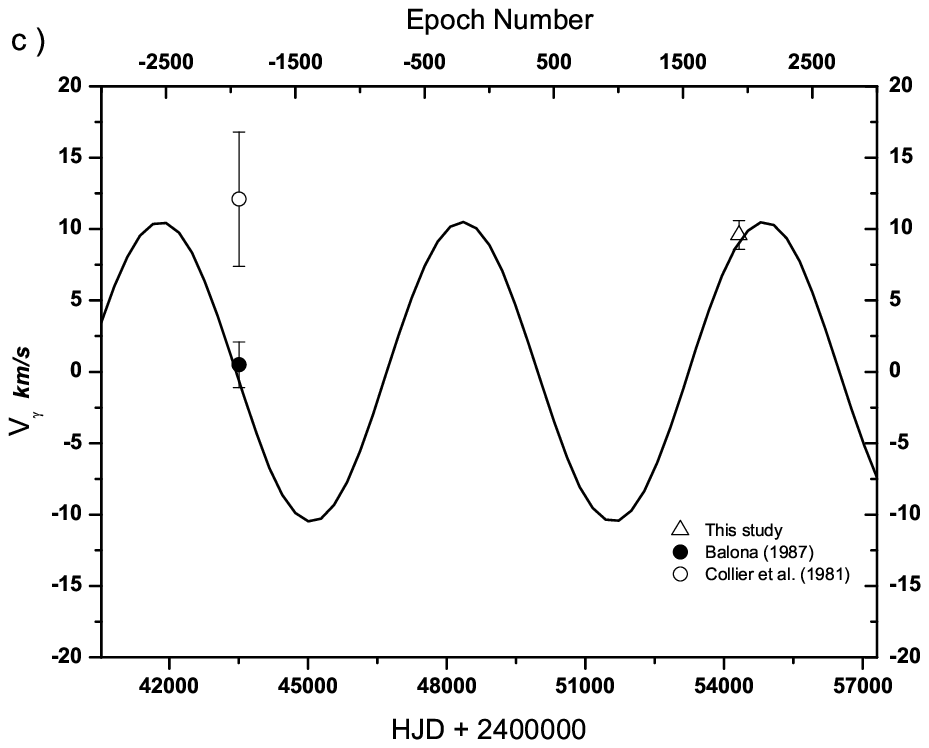}} &
      \resizebox{80mm}{!}{\includegraphics{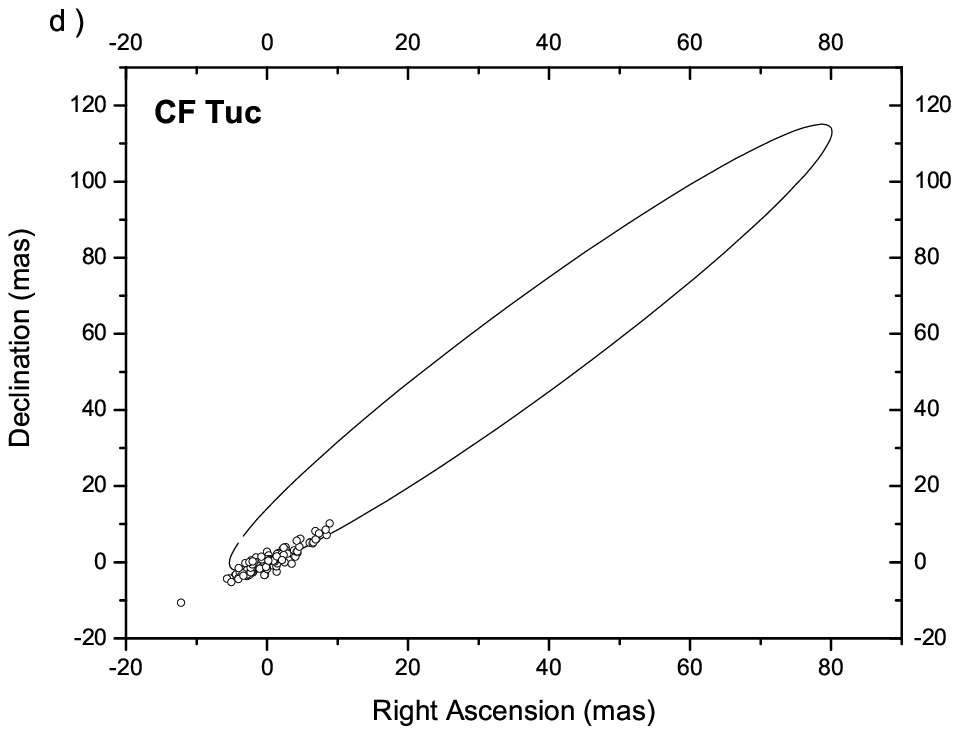}}
\end{tabular}
\caption{$(a)$ $O-C$ diagram of CF Tuc constructed with Eq.~(3). It
shows that the period variation of the system could be interpreted
in terms of two abrupt period changes. The residuals were calculated
separately with the three linear ephemerides given Eqs.~(4,5,6).
$(b)$ Sinusoidal representation (solid line) is superimposed on the
parabolic form (dashed line) of the $O-C$ variation of CF Tuc, and
the residuals from the best fit curve. The $O-C_{1}$ values in the
top panel were calculated by using the linear ephemeris of Anders et
al. (1999), while those of $O-C_{2}$ in the second panel were
computed by Eq.~(8) to show a symmetric parabola at the vertex with
$E$=0. These panels show the same parabola + sinusoidal variation.
$(c)$ Systemic velocity variations in CF Tuc. The individual points
show observed systemic velocities; the continuous curve is the
radial velocity curve corresponding to the light-time orbit in
Table~\ref{table7}. $(d)$ A relative astrometric orbit of CF Tuc about a
barycentre with a third body. The points represent Hipparcos
astrometric data, while the solid curve corresponds to the
astrometric solution given in Table~\ref{table7}.} \label{fig9}
\end{center}
\end{figure}

The $O-C$ values versus $E$ values (and years) were found and the
results are shown in the upper panel of Fig.~\ref{fig9}a,b.
\cite{t2}  first noted that the orbital period of the system changes
in the form of an upward parabola and tried to explain this
variation in terms of a mass transfer or a mass loss from the
system. \cite{a} showed that the orbital period change has a cyclic
character and discussed the $O-C$ diagram, the spot-wave amplitude
and the mean light change of the system as being due to the
Applegate mechanism. \cite{h5} and \cite{h6a} reported that the
orbital period of the system did not show any change between 1995
and 2006. Therefore, we could say that the real nature of the period
variation shows up as the data increases by time.

In the present $O-C$ analysis, due to a large scatter, one
spectroscopic time of minimum obtained by \cite{h2} was discarded
and altogether 33 photometric data were used. The standard errors of
observed minima times, as given by authors, are shown as error bars
in Fig.~\ref{fig9}a,b. The weights were assigned according to these
errors. As the standard errors are given in 4, 3 and 2 decimal
places, we used 10, 7 and 5 for weights, respectively. The observed
long--term period decrease of CF Tuc from these $O-C$ diagrams could
thus be explained as follows:

\emph{(i) Abrupt period changes:} The $O-C$ diagram in
Fig.~\ref{fig9}a was considered in terms of abrupt period changes.
Period jumps might have occurred two times within an interval of
about 30 years. Considering that the period has remained constant
between these two jumps, we calculated following three linear
ephemerides. The first ephemeris valid for $E$ $\leq$ 776 is:

\begin{equation}
HJD~(Min~I) = 2444219.2615(52) + 2^{\rm d}.797692(12) \times E,
\end{equation}
for 885 $\leq$ $E$ $\leq$ 1835 we have:
\begin{equation}
HJD~(Min~I) = 2444219.2126(51) + 2^{\rm d}.797749(4) \times E,
\end{equation}
and finally for $E$ $\geq$ 2192 we have:
\begin{equation}
HJD~(Min~I) = 2444219.7446(153) + 2^{\rm d}.797482(5) \times E.
\end{equation}
The first abrupt period change $\Delta P/P$=($-2.04\pm0.31$)$\times
10^{-5}$ occurred at HJD 2446527$\pm$70; while the second one
$\Delta P/P$=($+9.51\pm0.06$)$\times 10^{-5}$ occurred at HJD
2449745$\pm$100. The time interval between these two possible abrupt
period changes was taken as 3218$\pm$105 days (or approximately 9
years). Such sudden period jumps could be caused by anisotropic mass
ejections from one (or both) component(s) (e.g. \cite{h5h}). Indeed,
the RVs of H$\alpha$ emission support such a high velocity ejecting
gas from the active component (see Section~5), like a prominence in
the solar atmosphere. However, as seen in Fig.~\ref{fig9}a, the
sudden period changes seem to have occurred in the pattern of one
period increase and a subsequent decrease. This might be an
indication of sinusoidal variations rather than abrupt period
changes due to sudden mass ejections.

\emph{(ii) Continuous period change and the light-time travel
effect:} A reasonable fit to the $O-C$ data is obtained by using a
sinusoidal ephemeris with a quadratic term, as

\begin{equation}
C = T_{0} + P\times E + Q \times E^{2} +
A_{s}sin[\frac{2\pi}{P_{s}}(E-T_{s})],
\end{equation}
where $A_{s}$ is the semi-amplitude, $P_{s}$ the period and $T_{s}$
the time of minimum. However, to show the parabolic $O-C$ change
clearly, we recalculated $O-C$ values by using the following light
elements;

\begin{equation}
HJD~(Min~I) = 2448922.2310 + 2^{\rm d}.797630 \times E.
\end{equation}
These new $O-C_{2}$ values were then plotted against the epoch
number and observation years in the second panel of
Fig.~\ref{fig9}b. This panel shows a simple downward parabola, where
the axis of symmetry is parallel to the $O-C$--axis with the vertex
at epoch number $E$=0. It should be noted that the best model
fitting both $O-C_{1}$ and $O-C_{2}$ values yields the same values
within uncertainties for both quadratic and sinusoidal terms.
Finally, parameters of the best theoretical curve fitting the $O-C$
data are given in Table~\ref{table7}, and the best theoretical fit
with the observational data is plotted in Fig.~\ref{fig9}b.

\begin{table*}
   \caption{Parameters for the CF Tuc sinusoidal O-C solution + astrometry.}
\label{table7}
\begin{tabular}{ll}
\hline
Parameter & Value \\
\hline
\it Sinusoidal O-C solution & \\
$T_{0}$ (HJD) & 2448922.2102${\pm}$ 0.0021 \\
$P$ (d)     & 2.797641${\pm}$ 0.0000015 \\
$Q$ (d)     & $-4.9\times10^{-8}{\pm}2.1\times10^{-9}$ \\
$A_{s}$ (d) & 0.0363${\pm}$ 0.0018 \\
$P_{s}$ (yr)& 17.87${\pm}$ 0.57 \\
$T_{s}$ (HJD) & 2430387${\pm}$ 599 \\
$\Sigma W (O-C)^{2}$ & 0.00848 \\
\hline \it Astrometry solution & \\
$P_{12}$ (yr)& 17.87 \\
$T_{12}$ (HJD) & 2443441 \\
$a_{12}$ (mas) & 72.04 \\
$e_{12}$             & 0 \\
$\omega_{12}$ (deg)       & 90 \\
$i_{12}$ (deg)       & 83 ${\pm}$ 7 \\
$\Omega_{12}$ (deg)       & 144 ${\pm}$ 19 \\
$ \Delta \alpha cos \delta $ (mas)  & 37.4${\pm}$0.2  \\
$ \Delta \delta $ (mas)  & 56.4${\pm}$0.2  \\
$ \Delta \mu _{\alpha}cos \delta$ (mas/yr)  & -2.9${\pm}$0.3  \\
$ \Delta \mu _{\delta}$ (mas/yr)  & 0.7${\pm}$0.3  \\
$ \Delta \Pi$ (mas)  & -0.7${\pm}$0.3  \\
$\chi ^{2}/\nu$   & 1.01 \\
 \hline
\end{tabular}
\end{table*}

According to the quadratic term given in Table~\ref{table7}, the orbital period
of CF Tuc is continuously decreasing at a very rapid rate of
$1.11\pm0.05$ seconds per year. Here, CF Tuc appears to have the
highest rate of period decrease among the RS CVn systems. We
considered the combined effect of the mass loss and the mass
transfer to study this observed period decrease of the system and
used the following equation, given by \cite{e2a} and \cite{e2b}:

\begin{equation}
\frac{\Delta P}{P}=3\left( \frac{r_{A}}{a}\right) ^{2}\frac{\delta
M}{M}+3\frac{(M_{l}-M_{g})}{M_{l}M_{g}}\Delta M,
\end{equation}
where the mass $\Delta M$ is transferred from the mass--losing
component to the gainer, the $\delta M$ is the amount of mass lost
from the system after co--rotating with the system up to the
distance $r_{A}$ (i.e. Alfv\'{e}n radius), and $\Delta P/P$ is the
period change. Since CF Tuc is a detached system, in which the
primary and secondary components are filling $\sim52\%$ and
$\sim89\%$ of their lobes (see Section~6), a direct mass transfer
between components is not expected. However, the secondary component
is a magnetically active and larger star, which is not far from
filling its Roche lobe, and then there could be weak coronal flow
from this component to the primary component through the inner
Lagrangian point. Therefore, the active component might have a
strong stellar wind, which drives the mass loss and mass transfer in
the system. The RVs of $H\alpha$ emission support such a strong
stellar wind, which reaches twice the larger distance than the
radius of the active, subgiant component (see Section~5). If we
assume that the transferred mass due to the wind from the secondary
to the primary component and the co-rotating distance are $10^{-11}$
$M_{\sun}$/yr and $10R_{2}$, respectively, then Eq.~(9) gives the
mass loss rate of $\delta M$ = $3.38\times 10^{-7}$ $M_{\sun}$/yr
for the observed period change of $\Delta P/P$ = $-4.57 \times
10^{-6}$ yr$^{-1}$. It is worth mentioning that this mass loss rate
is 10 times higher than the maximum value of the range between
10$^{-11}$ and $10^{-8}$ $M_{\sun}$/yr given by \cite{h4} for the
mass losses due to winds from red-giant stars.

There are two plausible causes of the sinusoidal $O-C$ variation: a
light-time effect due to a third body in the system and a period
modulation due to the magnetic activity cycle of one of the components.
We shall investigate these suggested hypotheses in turn.

According to Table~\ref{table7}, CF Tuc would have a circular orbit
around the center of the mass of a three-body system and its period
would be $17.87\pm0.57$ yr. The projected distance of the center of
mass of the eclipsing binary to that of the three-body system would
be $6.29\pm0.31$ AU. These values lead to a large mass function of
$f(M_{3})$ = $0.78\pm0.07$ $M_{\sun}$ for the hypothetical third
body. The mass of such a third body would then range from
$9.63\pm0.29$ $M_{\sun}$ for $i_{12}$=30$^\circ$ to $2.71\pm0.13$
$M_{\sun}$ for $i_{12}$=90$^\circ$. Here the sum of masses was taken
as $M_{1}$ + $M_{2}$=2.34 $M_{\sun}$ (see Section~8). If the third
body were co--planar with the eclipsing pair, its mass and the
radius $r_{3}$ of its orbit around the center of mass of the
three-body system would be about $2.99\pm0.13$ $M_{\sun}$ and 4.93
AU, respectively. This value of $r_{3}$ is smaller than the radius
of the orbit of Jupiter, however, it shows that the third body would
revolve far beyond the outer Lagrangian points of CF Tuc, and its
orbit should be stable. If we consider the distance of CF Tuc as 89
pc (see Table~8), the minimum projected angular separation between
the third star and the eclipsing pair could be estimated as $\sim71$
mas.

The semi-amplitude of the radial velocity of the center of mass of
the eclipsing pair, relative to that of the three-body system, is
derived to be 10.5 km/s, which is a convenient value for modern
spectroscopic observations to resolve reliably. The theoretical
variation of the systemic velocity of CF Tuc, caused by the orbital
motion around the common barycentre, is illustrated in
Fig.~\ref{fig9}c. There are three values of the systemic velocity
observed at different epochs: 12.1$\pm$4.7 km/s (\cite{c3c}),
0.5$\pm$1.6 km/s (\cite{b00}), and 9.6$\pm$1 km/s (present study).
Except for the first data point, the observed systemic velocities
follow the long-term variation corresponding to the light-time orbit
in Table~\ref{table7} and are almost the same (within their standard
erros) as the theoretical values. The velocity measurement by
\cite{c3c} is affected by a relatively large error, which could be
caused by the scatter of RV data points and the two year time span
of their observations.

The astrometric method was also used to check the third body
hypothesis. A similar procedure was applied by \cite{r1b} for R CMa,
by \cite{b01} and \cite{b02} for XY Leo and $\delta$ Lib, by
\cite{b4b} for U Oph and also \cite{z1a} for VW Cep, $\zeta$ Phe and
HT Vir. We used the Hipparcos Intermediate Astrometric Data
(\cite{esa97}) for this. Hipparcos observed CF Tuc between January
1990 and January 1993. There are 76 one-dimensional astrometric
measurements corresponding to 40 different epochs in the Hipparcos
Intermediate Astrometric Data, which were obtained by the two
Hipparcos data reduction consortia: FAST and NDAC. These data are
available from the Hipparcos web page. In fact, the astrometric
method gives support to the third body hypothesis on orbital period
analyses of eclipsing binaries in two ways: one is to plot an orbit
of the eclipsing binary about the barycentre of a three-body system
and the other is to determine its orbital inclination ($i_{12}$). We
followed the procedure applied by \cite{b4b} for our target. In this
procedure, an orbital model, which is derived from the orbital
motion of the eclipsing binary around the barycentre of a three-body
system, is convolved with the astrometric motion (parallax and
proper motion). This model has 12 independent parameters: $a_{12}$,
$e_{12}$, $\omega_{12}$, $i_{12}$, $P_{12}$, $T_{12}$ (periastron
passage time), $\Omega_{12}$ (seven for the orbital parameters), and
$\alpha$, $\delta$, $\mu_{\alpha}$, $\mu_{\delta}$, $\Pi$ (five for
the astrometric components; equatorial coordinates + proper motion +
parallax). Since the Hipparcos astrometric data cover only 1/6 part
of the orbital period of the hypothetical three-body system, we
could take only two parameters from the orbital parameters, $i_{12}$
and $\Omega_{12}$, together with five adjustable astrometric
parameters. The final results are given in Table~\ref{table7} and
Fig.~\ref{fig9}d. The orbital inclination of CF Tuc in the triple
system is, $i_{12}$, about 83$^\circ$ and gives the mass of a third
body as about 2.74 $M_{\sun}$. Unfortunately, since the time span of
the Hipparcos observations is much shorter than the orbital period
of the three-body system and is not the periastron passage of CF Tuc
in the three-body orbit, the data, as shown in Fig.~\ref{fig9}d,
cover only a small part of the orbit.

An alternative way of explaining the O-C behavior would be
Applegate's mechanism (\cite{apple92}). According to this, the
cyclic magnetic activity could produce orbital period modulations,
which are observed in some eclipsing binaries, especially in RS CVn
type systems. Magnetic activity can change the quadrupole moment of
a component as the star goes through its activity cycle. The cyclic
exchange of angular momentum between the inner and outer parts of
the star can change both the shape and radial differential rotation
of the star. The torque required for such transfer of angular
momentum could be provided by a subsurface magnetic field of several
kG. Any change in the rotational regime of a component of an
eclipsing binary due to magnetic activity could be reflected in the
orbit, as a consequence of the spin-orbit coupling. Here, we shall
use Applegate's formalization to examine the sinusoidal part of the
orbital period variation of  CF Tuc and assume that the secondary
star could be responsible for the observed orbital period
modulation. The $O-C$ diagram of CF Tuc, given in Fig.~\ref{fig9}b,
shows a modulation with a semi-amplitude of 0.0363 days and a
modulation period of 17.87 years. This gives $\Delta
P/P$=3.49$\times$10$^{-5}$. The change in the orbital period is
$\Delta P$=8.45 s. The angular momentum transfer would be $\Delta
J$=6.84$\times10^{48}$ gcm$^{2}$s$^{-1}$. If the mass of the shell
is $M_{s}$=0.1$M$, the moment of inertia of the shell is
$I_{s}$=1.03$\times10^{55}$ gcm$^{2}$, and the variable part of the
differential rotation is $\Delta\Omega/\Omega$=0.026. The energy
budget and RMS luminosity variation are $\Delta
E$=9.08$\times10^{42}$ ergs and $\Delta L_{RMS}$=13.23 $L_{\sun}$.
This model gives a mean subsurface field of 12.1 kG. The RMS
luminosity variation predicted by this  model is larger than the
total luminosity of the active star. Therefore, a model with
$M_{s}$=0.1$M$ and $\Omega_{dr}$=$\Delta\Omega$ cannot explain the
orbital period change observed in CF Tuc. \cite{apple92} has also
calculated a similar result for RS CVn itself. However, he could
obtain a reasonable result for RS CVn using the following two
approaches: one is that the active component has solid body
rotation, in which $\Omega_{dr}$=0. The other is energy dissipation
in the inner part of the star due to differential rotation and some
storage of energy in the convection zone. This energy could be
omitted from the luminosity variation. These two modifications lower
the luminosity variation by a factor of 4 but we are still left with
a large value of $\Delta L_{RMS}$=3.31 $L_{\sun}$=0.85 $L$.
Therefore, we conclude that the Applegate mechanism is not
sufficient to explain the observed period changes.

\section{Results and discussion}

The new 24 high-resolution \'{e}chelle spectra of CF Tuc were
analysed and precise spectroscopic orbital elements were obtained by
means of two techniques; cross-correlation and spectral
disentangling. The KOREL program was applied to four spectral
orders, which contain about 15 lines, and then reliable radial
velocities of both components of the system were obtained. In the
literature, there are two spectroscopic studies on CF Tuc:
\cite{c3c} and \cite{b00}. Authors of the former  obtained 31
spectra of the system between 1976 and 1978 and used H$_{\delta}$
absorption and CaII H and K emission lines to derive radial
velocities of both components. However, their data have a large
standard error of 14 km/s, and their spectroscopic orbital elements
were less accurate than these presented in this work. \cite{b00}
gave radial velocity measurements of only the hotter component and
its orbital solution.

Using the observed spectral lines with high precision (i.e. Na D2
line), we found the components of CF Tuc to be in synchronous
rotation. In fact, the derived rotational velocity of this secondary
star is puzzling, a point which was examined by \cite{c3b}. They
noted that there is a discrepancy between measurements of the
rotational velocity of the secondary published in the literature
(\cite{b3}; \cite{a}; \cite{d2}). The rotational velocity of the
primary (hotter component) ranges from 25 to 30 km/s, which is
similar within uncertainties, while that of the secondary component
is 52 to 70 km/s, which corresponds to a rather large range of
values in the secondary star radius, from 3 $R_{\odot}$ to 4.3
$R_{\odot}$. \cite{c3b} used the measurements of \cite{d2} and
emphasized that the projected rotational velocity of the secondary
should be $\sim70$ km/s from the derived absolute parameters of the
system if the components rotated synchronously.

Our radial velocity and the 2007 $BV$ light curves from Innis, were
simultaneously solved using Wilson-Devinney code. The 2007 $BV$
light curves show large asymmetry in the two different maxima
similar to those observed in RS CVn type eclipsing binary stars. CF
Tuc is a bright system ($V\cong7.6$ mag), and has been frequently
observed photometrically. In the last three decades about 30 light
curves of the system were obtained; almost all of them exhibit large
asymmetries. The secondary component is apparently magnetically
active, as Collier et al. (1981) observed the indicator of magnetic
activity, CaII H \& K emission lines, from this component. Our
H$\alpha$ and CaII H \& K observations also show that the secondary
component is a chromospherically (or magnetically) active star.
Therefore, these light asymmetries were considered as maculation
effects and interpreted using spot models on the secondary.
\cite{b4a}  solved 25 light curves, mainly  taken in broadband $V$
and spanning a 16 year period, and modelled the spot activity of the
secondary (cooler) component using their program (ILOT). They
suggested that the spot luminosity  has been decreasing over the 16
year period. \cite{a} solved 27 light curves of the system, taken
between 1979 and 1996, and estimated the parameters of the spot
placed on the secondary, especially the longitudes and radius of the
spots. They suggested that there was a strong tendency for spots to
appear in a narrow range of longitudes, just before the phase of
primary minimum and just after secondary minimum. \cite{a}
determined than in the 2007 $BV$ light curves, the maculation wave
begins to appear just after the phase of secondary minimum.
Therefore, we used one dark spot on the secondary to solve these
light curves. On the other hand, the Hipparcos light curve shows two
distortion waves; the first one appears just after the secondary
minimum like the 2007 light curves and the other is in the primary
minima. Therefore, we introduced two dark spots on the secondary
component to solve  the Hipparcos light curve. Since the parameters
obtained from the simultaneous solution of $BV$ light and RV curves
are more reliable, during the iterations of the Hipparcos light
curve only spot parameters and luminosity of the primary have been
adjusted. A comparison of the light curve asymmetries with H$\alpha$
emission modulation suggests a close spatial association between the
photospheric spot(s) and chromospheric/coronal active region of the
secondary component. Such associations have already been found in
several CAB systems (e.g. \cite{fb0}; \cite{c0}; \cite{b3a};
\cite{fb}).

The simultaneous solution of $BV$ light and radial velocity curves
allows us to calculate the absolute parameters of CF Tuc. The
resulting parameters (with uncertainties) are given in
Table~\ref{table8}. As mentioned above, following Anders et al.
(1999) and Budding and McLaughlin (1987) we adopted an effective
temperature of about 6100 K for the primary component of the system.
Anders et al. found it to be in good agreement with the value
derived from the colour indices of Collier (1982), and Budding and
McLaughlin estimated it using the spectral type of the primary
revised by Collier et al. (1982). If the standard error of the
photometric observations by Collier is taken as 0.01 mag, this
corresponds to an uncertainty in the primary component's temperature
of 200 K. On the other hand, the uncertainty in the temperature of
the secondary component, 19 K, given in Table~6, is the formal
$1\sigma$ error coming from the WD simultaneous solution. The
corrected uncertainty could be estimated as 219 K based on the
uncertainty of 200 K in the effective temperature of the primary. In
the calculations, the temperature, bolometric magnitude and
bolometric correction of the Sun were taken as 5780 K, 4.75 mag and
-0.14 mag, respectively. Bolometric corrections for the components
of the system were taken from the tables of \cite{z2}. From the
distance modulus of 2.87$\pm$0.15 mag, we derived the distance of
the system to be 89$\pm$6 pc, under the assumption of $A_{V}$=0.
According to the new Hipparcos parallax given by \cite{lb}, the
distance to CF Tuc is about 89$\pm$4 pc. This consistency between
the dynamic and Hipparcos parallaxes shows the accuracy of the
determined absolute parameters of CF Tuc.

\begin{table}
  \caption{Absolute parameters of the CAB star CF Tuc} \label{table8}
\begin{tabular}{lcc}
\hline
\textbf{Parameter} & \textbf{Primary} &  \textbf{Secondary} \\
\hline
$M$ ($M_{\odot}$) & 1.11${\pm}$0.01    & 1.23${\pm}$0.01 \\
$R$ ($R_{\odot}$) & 1.63${\pm}$0.02    & 3.60${\pm}$0.02 \\
Log $g$ (cgs)     & 4.05${\pm}$0.02    & 3.42${\pm}$0.02 \\
$T$ (K)         & 6100${\pm}$200       & 4286${\pm}$219 \\
$M_{bol}$      & 3.45${\pm}$0.17    & 3.27${\pm}$0.23 \\
$L$ ($L_{\odot}$) & 3.32${\pm}$0.51    & 3.91${\pm}$0.84 \\
$M_{V}$        & 3.50${\pm}$0.17    & 3.80${\pm}$0.23 \\
$M_{bol}$ $(system)$ & \multicolumn{2}{c}{2.60$\pm$0.15} \\
$M_{V}$ $(system)$ &  \multicolumn{2}{c}{2.87$\pm$0.15} \\
$d$ (pc)         & \multicolumn{2}{c}{89${\pm}$6}  \\
\hline
\end{tabular}
\end{table}

\begin{figure}
\begin{center}
\begin{tabular}{cc}
      \resizebox{80mm}{!}{\includegraphics{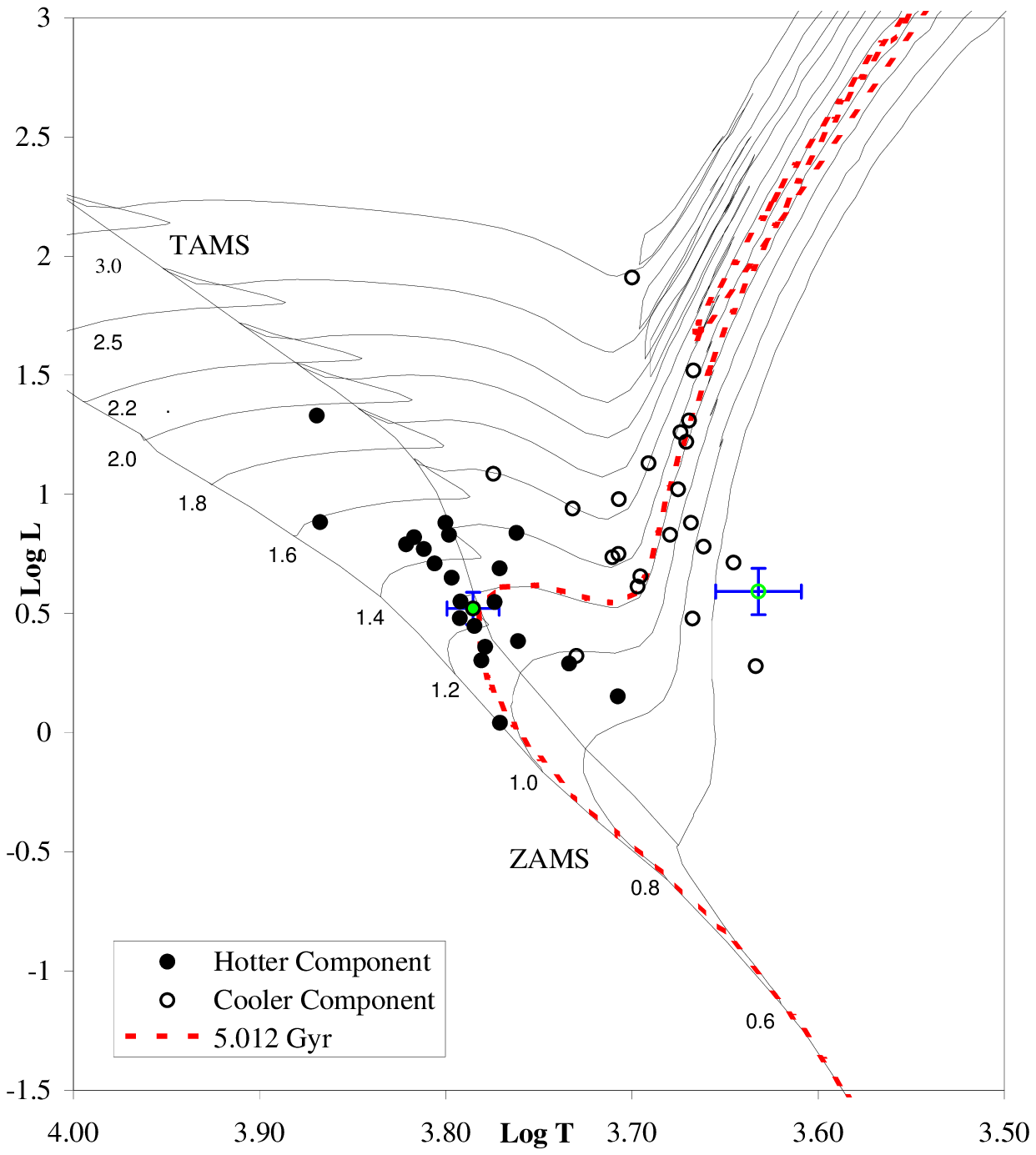}} &
      \resizebox{80mm}{!}{\includegraphics{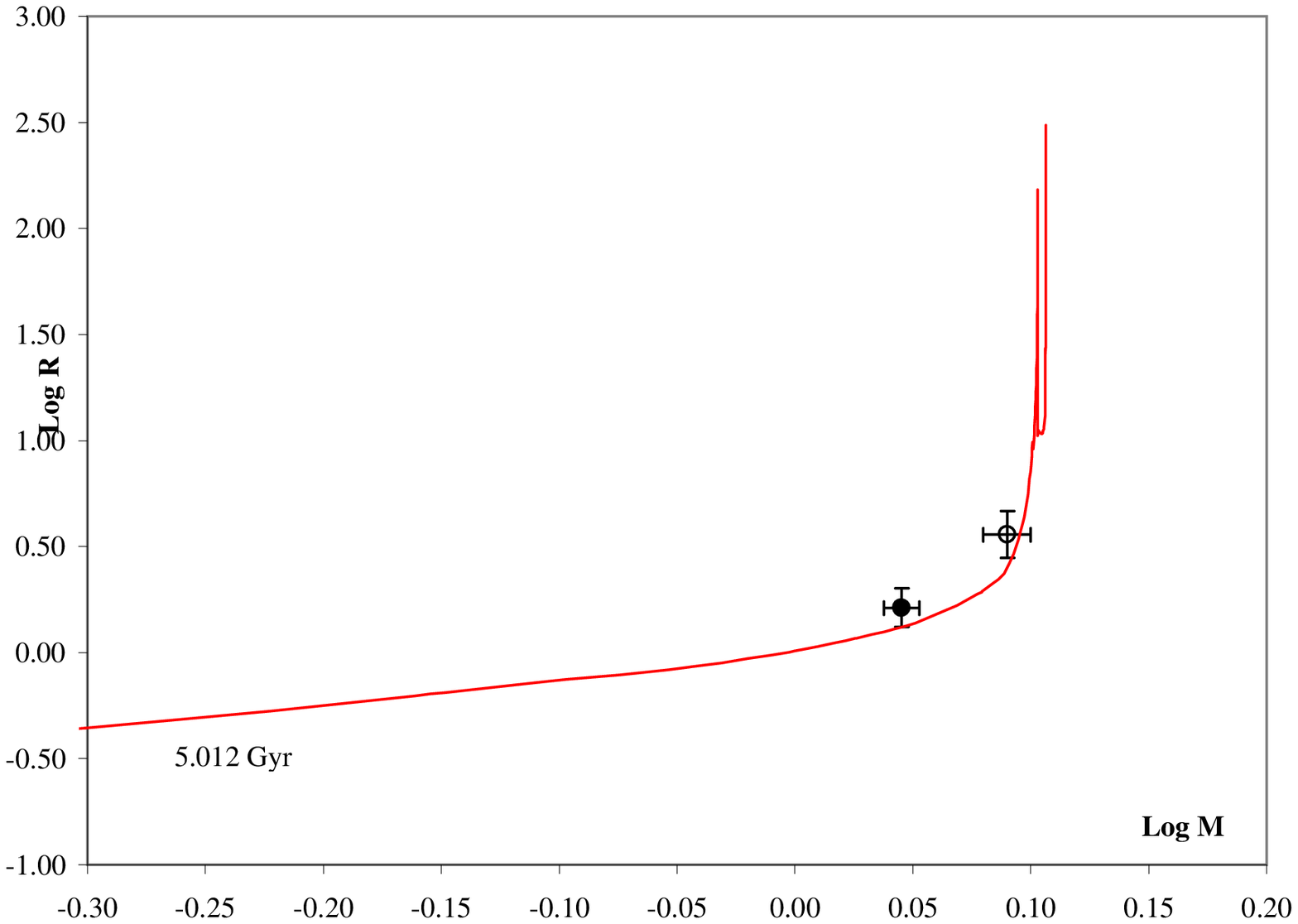}}
\end{tabular}
\caption{Hotter and cooler components of CF Tuc together with those
of the other RS CVn type eclipsing binaries, in which the cooler
component is more massive (data from Eker et al. 2008), plotted in
the HR diagram (left panel). Comparison between evolutionary models
and the physical parameters of CF Tuc in the mass-radius diagram
(right panel). The filled and open circle symbols represent the
hotter and cooler components in the binaries, respectively. The
error bars of the measured quantities are shown by vertical and
horizontal lines. Zero Age Main Sequence (ZAMS), Terminal Age Main
Sequence (TAMS), the evolutionary tracks and isochrone were taken
from Girardi et al. (2000) for the solar chemical composition. The
numbers denote initial masses.} \label{fig10}
\end{center}
\end{figure}

The locations of the components of CF Tuc in the
luminosity-effective temperature ($T_{eff}$--$L$, i.e.
Hertzsprung-Russell diagram) and the mass-radius ($M$--$R$) planes
are shown in Fig.~\ref{fig10}. We considered only RS CVn type
eclipsing binaries, in which the cooler component is a more massive
one, to compare the CF Tuc system with other RS CVn's. According to
these diagrams, while the cooler component has evolved behind the
terminal age Main Sequence, the hotter one is still on the Main
Sequence, approaching the TAMS. This case predicates that CF Tuc is
similar to other RS CVn's. We compared our observationally
determined physical parameters with those inferred from the
evolutionary tracks in the HR diagram. The best fit appears at age
of 5.012 Gyr for both components. The PARAM program at web page
(http://stev.oapd.inaf.it/~lgirardi/cgi-bin/param) based on the
Bayesian method of da Silva et al. (2006) gives   the CF Tuc age as
5.3 Gyr, which agrees well with our determination.

The orbital period change of the system is interesting and difficult
to solve. The $O-C$ data of CF Tuc could be represented either by
two abrupt period changes or by a sinusoidal period variation
superimposed on a downward parabola. In the first case, there
appears to be two sudden period jumps: in 1986 and 1995. The first
jump shows a period increase, while the second denotes a period
decrease. According to the conservative mass transfer mode, the
period increase corresponds to a mass transfer from the less massive
to the more massive component. In the case of CF Tuc, since the less
massive (primary) component is quite far from filling its Roche
lobe, this period increase derived from the first jump can not be
explained by the mass transfer mechanism. If we take into account
the representation of $O-C$ sinusoidal variation superimposed on a
parabola; the downward parabola indicates the highest rate of period
decrease among the CAB systems. However, since CF Tuc is a detached
system, we do not expect direct mass transfer between the
components. Therefore, another  possible  explanation could be a
mass loss by stellar wind from the subgiant component. Since the
active, subgiant component fills $\sim89\%$ of its Roche lobe, if a
large amount of escaping mass by stellar wind transfers to the
primary one, the remaining amount of escaping mass would leave the
system. To check this hypothesis, the H$\alpha$ and CaII H \& K
spectra of the system were examined. Since the RVs of H$\alpha$ and
CaII H \& K emission features follow the orbital motion of the
secondary (cooler) component, these emission features should come
from chromospheric and/or coronal layers of that component. From the
large width of the H$\alpha$ emission line profile, we estimate
turbulent velocities to be up to 200 kms$^{-1}$. Such velocities
well support our hypothesis that the emission features originate in
the circumstellar material.

The sinusoidal form of the orbital period variation was considered
as an apparent change and interpreted in terms of the light-time
effect due to an unseen component in the system. The large amplitude
($\sim$0.04 day) and small period ($\sim$18 years) of the sinusoidal
form of the $O-C$ diagram give quite a large value (2.7 $M_{\sun}$)
for the minimum mass of the hypothetical third body. The observed
systemic velocity variation and Hipparcos intermediate astrometric
data of CF Tuc partially support the hypothesis of the existence of
a third body in the system. If such a third body were a
main-sequence star, it would be a blue dwarf of a late-B spectral
type. However, neither new high-resolution spectroscopic
observations nor photometric analysis show any evidence to confirm
the presence of such a star.  Additionally, we could not expect such
a young star as the binary system is  much older. Therefore, the
hypothetical third body must make a negligible optical contribution
to the total light. With this mass, it could be either a massive
neutron star (NS) or a black hole (BH). In the case of the companion
being a compact object, the question is, can we see it, i.e. in
X-rays? CF Tuc was observed by \cite{franciosini} and they detected
X-ray photons with energies of one keV in quiescence and a few keV
during the flares. This soft X-ray emission was identified as being
released in the corona of the magnetically active component. A young
NS should give much higher X-ray emission, and an isolated NS would
produce a termal X-ray emission of 40-100 keV (\cite{trumper}). The
progenitor of the NS must be a B-type star which evolves fast
through the Main Sequence, within 100-500 million years, depending
on the mass. The newborn NS would have a temperature of 10$^{7-8}$~K
and could be detected in X-rays. However, assuming the triple system
has been formed at the same time and not via a capture, the NS would
have enough time to cool down to a much lower temperature. According
to the NS models, after 10$^{7}$ years, the NS temperature would be
only 10$^{5}$~K. The other possibility, if either a NS or a BH is
the third component, could be the accretion luminosity when matter
is being accreted onto this object. In fact, a black hole can be
seen only through its accretion effects. Due to the large distance
from the binary system to the companion (4.9AU), the only
possibility for accretion would be through a stellar wind from the
binary. Even though CF Tuc has the higher rate of stellar wind among
CAB binaries, at the third body separation only a tiny fraction of
mass lost by the active component could be accreted, as such
accretion is much less efficient than that through the Lagrangian
point. We shall estimate the rate according to the formula given by
\cite{frank}. Taking $\delta M$ = $3.38\times 10^{-7}$
$M_{\sun}$/yr, separation of 4.93 AU, mass of the accreting object
to be 3 $M_{\sun}$ and derived radius and mass of the secondary
star, we calculated the accretion luminosity to be about
$L_{acc}=3.4\times 10^{34}$~erg/s. Such a value is comparable to the
quiescent luminosities of NSXN and BHXN (\cite{narayan}) and
quiescent low-mass X-ray binary transients (\cite{lasota}). If the
third companion is a NS with such a luminosity it should be observed
in X-rays, unless the efficiency of the mass transfer to the primary
is more efficient than we have assumed. If the companion is a BH,
which has no surface, a significant fraction of mass could be lost
below the event horizon, instead of being converted into hard
photons.  Also, if the accretion onto the BH is spherical and not
through a bow shock, as considered above, it could be less efficient
as well. However, the factors describing this efficiency are within
a huge range, between $10^{-8}$ to $10^{-1}$ (\cite{frank},
\cite{shapiro1}, \cite{shapiro2}, \cite{shapiro4}) and strongly
model dependent (i.e. rotation of the BH, magnetic field, speed of
the BH or the accreted material). Concerning the efficiency of the
accreting matter being converted into photons, this factor could be
in the same range. If the factors are in the lower end, then the
existence of a BH as the third companion in CF Tuc in not
impossible, though unlikely, as the possibility of its observation
may fall below the detection limit and the light time effect would
be the only evidence of its existence.

Another explanation of the orbital period  sinusoidal variation
could be the Applegate mechanism. However, the Applegate model with
$M_{s}$=0.1$M$, $\Omega_{dr}=\Delta\Omega$, and $\Delta
L_{RMS}=0.1L$ cannot explain the rate of orbital period change
observed in CF Tuc.

The last possibility, we can consider, would be a spurious sine term
which may not repeat in the future. The  observed light curves have
strong asymmetries which  can affect accuracy of the minima times
determination. In some cases, these asymmetries are so large that
secondary minima can not even be easily recognized. Therefore, some
spurious $O-C$ residuals may come from the asymmetries of the light
curve. For instance, \cite{b4a} found an approximate 6 year magnetic
cycle, considering that these light curve asymmetries were caused by
maculation effects. Therefore, we conclude that only future
observations can reveal the true nature of the observed period
changes. If they are caused by a companion BH, this will be the
closest black hole to the Earth.

\section{Acknowledgements}
This study is related to the Southern Binary Project of the
Astrophysics Research Centre at \c{C}anakkale Onsekiz Mart
University, the Carter National Observatory, and the Mt John
University Observatory in New Zealand. It also forms part of the PhD
thesis of D. Do\u{g}ru and was supported by \c{C}anakkale Onsekiz
Mart University Research Foundation under grant no. 2007/55. We
thank Prof. John Hearnshaw for granting use of the observing
facilities at Mt John University Observatory. We also thank J. L.
Innis for making his data available. D. Do\u{g}ru thanks Professor
Edwin Budding and Dr. Hasan Ak for their useful comments and
suggestions. Discussions with M. Balucinska-Church and J.-P. Lasota
were greatly appreciated. We also thank the anonymous referee for
the comments which helped us to improve the quality of this paper.

\label{lastpage}

\end{document}